%% file: mainNOXPOL.tex
\documentclass[%
 reprint,
groupedaddress,
 amsmath,amssymb,
 aps,
]{revtex4-2}

\usepackage{graphicx}% Include figure files
\usepackage{dcolumn}% Align table columns on decimal point
\usepackage{bm}
\usepackage{xcolor}
\usepackage{subfigure}

\begin{document}

%\preprint{APS/123-QED}

\title{Time-Controlled Resonances in 2-D Metasurfaces via Equivalent Circuits}% Force line breaks with \\
%\thanks{A footnote to the article title}%

\author{J. Rafael Sánchez-Martínez$^{1}$}
\author{Mario Pérez-Escribano$^{2}$}
\author{\\Antonio Alex-Amor$^{3}$}
\author{Juan F. Valenzuela-Valdés$^{1}$}
\author{Carlos Molero$^{4}$}

\affiliation{$^{1}$Dpto. de Ingeniería de Comunicaciones, TELMA, Universidad de Málaga, Spain.}
\affiliation{$^{2}$Dpto. de Ingeniería de Comunicaciones, Telecommunication Research Institute (TELMA),\\
Universidad de Málaga, E.T.S. Ingeniería de Telecomunicación, 29010 Málaga, Spain.
}
\affiliation{$^{3}$Department of Electronic and Communication Technology, RFCAS Research Group,  \\ Universidad Autónoma de Madrid, 28049 Madrid, Spain.
}
\affiliation{$^{4}$Department of Electronic and Electromagnetism, Faculty of Physics, \\ University of Seville, 41012 Seville, Spain.
}

\newcommand{\red}[1]{\textcolor{red}{#1}}

\begin{abstract}
This work introduces a semi‑analytical frequency‑domain framework for the analysis of two‑dimensional, time‑modulated (2+1)‑D metasurfaces controlled by PIN diodes. The formulation focuses on the unit‑cell level, modeled as a waveguide discontinuity problem, where the space-time periodicity of the structure enables the representation of scattered fields via Floquet expansions. After appropriate mathematical treatment, these expansions lead to an equivalent circuit description of the metasurface, providing physical insight into its spatiotemporal scattering behavior and facilitating the design of reconfigurable electromagnetic devices. The model is employed to explore key phenomena present in space-time systems, such as frequency mixing and spatiotemporal scattering. In addition, dynamic tuning is explored in resonant metasurfaces, where time becomes an additional degree of freedom for the design. The dynamic control of resonances opens a new way to explore multi-band and wideband behaviors from very thin metasurfaces under temporal coupling.    

\end{abstract}

\maketitle

\section{\label{sec:Intro} Introduction} 

Over the past decade, the study of time-modulated systems in microwaves, millimeter waves, and photonics has experienced significant growth, driven by their disruptive and transformative properties \cite{Caloz2020_1, Caloz2020_2, Galiffi2022, Ptitcyn2023, engheta2023four}. These systems enable functionalities that are difficult to achieve with conventional time-invariant devices without relying on complex external circuitry. Notable examples include the realization of non-reciprocity \cite{Ramaccia2018, Wang2018, Prudencio2023, Silveirinha2023}, the generation of net gain \cite{Fleury2018, Pendry2021, Lyubarov2022, gaxiola2023growing, Feinberg2025}, and the manipulation of operating frequencies \cite{Xiao11, Yi2017, Liberal2023, Pacheco2025}. 

Although the underlying concepts were partly explored in pioneering studies during the 1960s and 1970s \cite{Morgenthaler1958, Simon1960, Oliner1963, Felsen70, Fante71, Fante73}, the practical implementation of modern time-modulated devices has only recently become feasible thanks to technological advances in the field \cite{Kord18circulator, wu2019isolating, Grbic2020, wu2022sideband}. In photonics and optics, reconfigurable materials such as graphene or indium tin oxide (ITO)-based compounds enable dynamic control of electromagnetic properties in the time domain \cite{CapassoGraphene2014, Tirole2023SlitTime, PangITO2021}. In contrast, in the RF and microwave regimes, electronic components such as PIN diodes and varactors are typically employed \cite{Grbic2020, CuiBreaking2019, CuiMapping2025}. Varactors offer the advantage of supporting a quasi-continuous range of states, enabling a wide variety of modulation profiles. PIN diodes, on the other hand, usually operate between two discrete states (ON and OFF), which greatly simplifies their biasing and control circuitry. To achieve even faster modulations, researchers have also explored devices based on photodiodes triggered by laser pulses, which can switch an order of magnitude faster than conventional PIN diodes~\cite{JonesTimeReflection2024}.

In particular, structured platforms,  periodic or quasi-periodic spatial arrangements of repeated individual elements, have greatly benefited from the inclusion of time modulations. The so-called time-modulated or space-time metamaterials have shown enhanced and more versatile diffraction behavior, including frequency conversion, nonreciprocal wave propagation, and tunable scattering patterns, compared to their time-invariant counterparts \cite{ZhanCui2018, Taravati2019, Tiukuvaara2021, Alex2022TV, Salva2024}. This is of great interest from an engineering perspective, as it enables the production of radiofrequency (RF) and optical metadevices with advanced functionalities that can be applied in modern dynamic communications and radar scenarios. 

Unfortunately, although research on space-time metadevices is rapidly advancing, commercial software capable of supporting their analysis and design remains limited. As a consequence, many of the physical phenomena associated with space-time modulation remain largely inaccessible to the broader research community. To the best of our knowledge, COMSOL Multiphysics is currently one of the few commercial platforms that allows the incorporation of time modulation alongside advanced metadevice configurations. However, as a general-purpose multiphysics tool, it can become cumbersome to use and yield inaccurate or highly inefficient simulations if the electromagnetic parameters and boundary conditions are not carefully defined. 

Consequently, many researchers have developed their own in-house computational tools to study time-modulated metadevices and time crystals. These approaches are typically based on analytical Floquet–Bloch formulations in the frequency domain \cite{Zurita2009, Tiukuvaara2021, wu2019isolating, Moreno2023}, matrix formalisms \cite{Ramaccia2021, EnghetaGaldi2021, MoleroABCD25}, or numerical finite-difference time-domain (FDTD) techniques \cite{Vahabzadeh2018, Stewart2018, ZapataTV2023, TaravatiEleftheriadesFDTD2025}.

In this context, we have introduced in our previous works a frequency-domain method for the analysis of periodic space-time metasurfaces composed of metallic elements \cite{Alex2022TV, Moreno2023, Salva2024}. The approach builds upon the analytical circuit models developed in \cite{Mesa2018, rodriguez2015analytical, 3D_Antonio, PerezEscribano_Lumped2024} for time-invariant metasurfaces, combined with Floquet–Bloch expansions of time-varying electromagnetic fields. The resulting analytical framework provides accurate predictions in frequency ranges where transient effects can be neglected, and highly conductive metals can be approximated as perfect electric conductors (PEC); namely, from DC up to the end of the millimeter-wave band \cite{ZapataTV2024}.

Our previous efforts have focused on describing the physics and engineering explorations of one-dimensional (1$+$1)-D time-modulated platforms. While these studies have provided valuable insights into fundamental phenomena such as frequency conversion and wave propagation, extending time modulation to higher-dimensional systems significantly expands the available physical mechanisms. In particular, two-dimensional time-modulated (2$+$1)-D platforms introduce an additional spatial degree of freedom that enables more complex resonant interactions involving transverse momentum components, which are not easily achievable in conventional (1$+$1)-D configurations. These additional degrees of freedom allow the excitation and coupling of richer modal spectra and can give rise to \emph{novel resonance mechanisms} associated with multidirectional wave propagation. Furthermore, the vectorial nature of electromagnetic fields becomes more fully accessible in (2$+$1)-D systems, opening the possibility of cross-polarization coupling and polarization conversion processes that are inherently absent or severely limited in purely one-dimensional structures \cite{CrosspolCircuit2021}. Consequently, exploring time-modulated platforms in higher dimensions offers a substantially broader design space for manipulating electromagnetic waves.

In this paper, we develop a frequency-domain formulation for the analysis of two-dimensional time-modulated (2$+$1)-D metasurfaces. Specifically, we focus on space-time metasurfaces composed of metallic elements with slot insertions, whose electromagnetic properties can be dynamically reconfigured via externally driven diode modulation, as shown in Fig.~\ref{fig:general}. The proposed formulation combines analytical circuit techniques, Floquet–Bloch expansions of the electromagnetic fields, and physical insight derived from external full-wave simulations. 

In contrast to our previous work, which described the system from a state-based perspective (e.g., transparent, reflective), the present model explicitly incorporates the unit-cell geometry. Consequently, the scatterer geometry becomes a central element of the analysis, along with the integration of reconfigurable components, such as PIN diodes, that enable temporal modulation of the metasurface response. In this framework, the scatterer's geometry and the time-modulation parameters jointly determine the electromagnetic behavior of the space–time metasurface. Importantly, this approach enables the resonant features of the structure to be dynamically tuned on demand, providing a powerful degree of control over the device response. Such tunable resonances are expected to play a crucial role in the engineering of adaptive electromagnetic platforms, with potential applications in advanced communication systems, dynamic filtering, and wavefront control.

%-------------------------------------
\section{\label{sec:Sect2} Theoretical framework}

\begin{figure}[!t]
  \centering
    \includegraphics[width= 1.02\columnwidth]{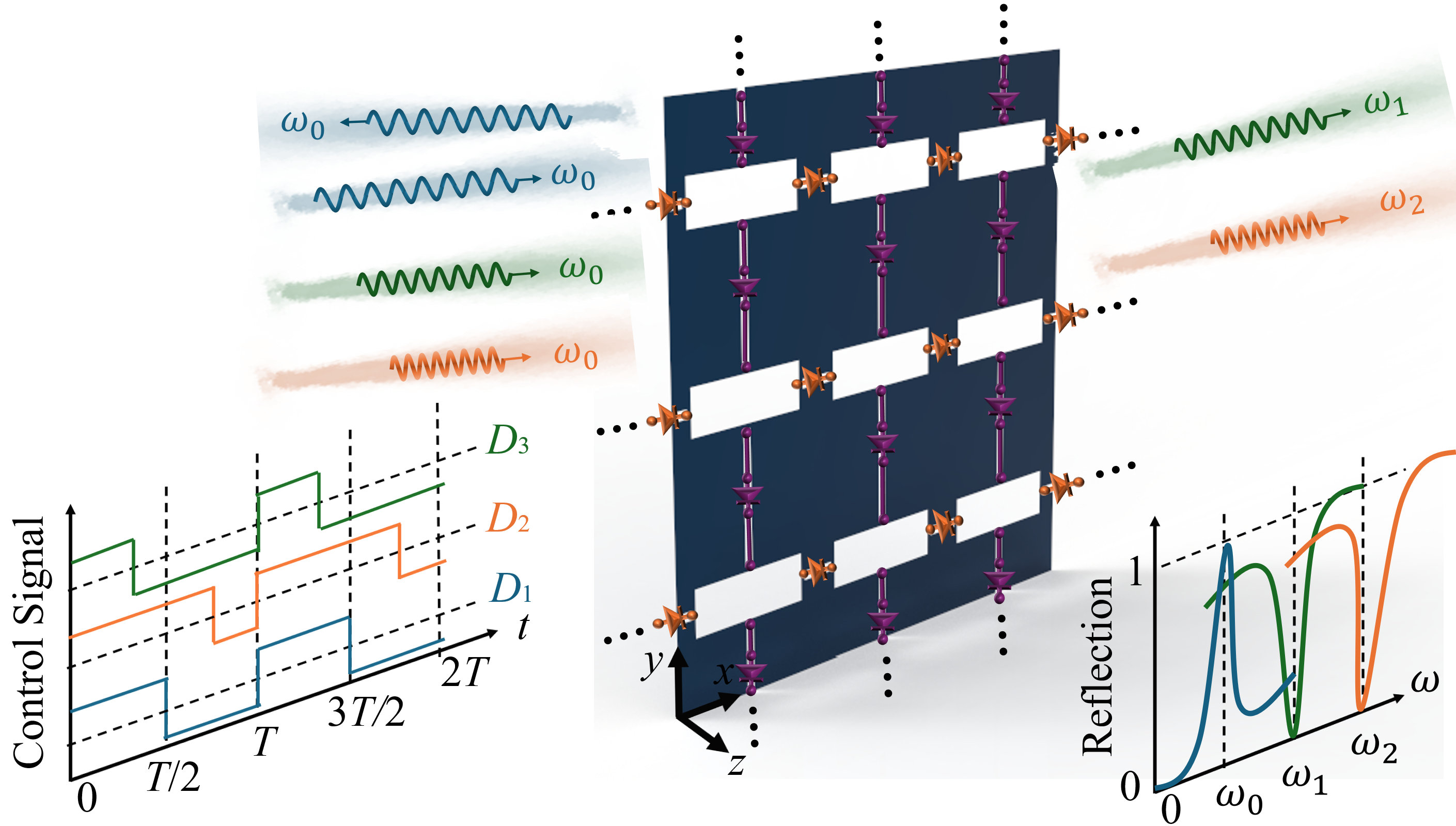}
  \caption{Sketch of a time-modulated reconfigurable metasurface. Diodes, which are biased via temporal control signals, act as the reconfigurable elements here. The shape of the control signals modifies the reflection/transmission properties of the space-time metasurface, enabling advanced additional functionalities such as frequency conversion and beam steering.  }
  \label{fig:general}
\end{figure}

The three-dimensional (2$+$1)-D problem, described by a generalized unit cell with two periodic spatial and one temporal dimensions, can be analyzed using equivalent circuit models. The approach adopted here draws on part of the principles reported in \cite{Berral2015}, originally developed for spatially periodic structures, and extends them by introducing time as an additional periodic variable.

Fig.~\ref{fig:general} illustrates the considered space–time metasurface, formed by a periodic array of time-modulated scatterers located in the XY plane and assumed to have negligible thickness ($d \ll \lambda_0$). In RF and microwave implementations, time modulation can be practically achieved using diodes externally biased with time-varying control signals. In the figure, $x$- and $y$-oriented diodes are shown with different colors to indicate that independent control signals may drive them. By adjusting the waveform parameters—such as the temporal period $T$, duty cycle $D$, and amplitude—it is possible to dynamically tune the electromagnetic response of the metasurface, enabling functionalities including controlled reflection and transmission, frequency conversion, and beam steering.

%-----------------------------------------------------------
\subsection{\label{sec:Sect2_1} Main remarks for the circuit-model derivation}

As depicted in Fig.~\ref{fig:general}, the structure is excited by a plane wave. The region where the wave impinges and is partially reflected is denoted as region (1), while region (2) corresponds to the transmission side. The fields in these regions are determined by the field distribution at the plane of the periodic structure, referred to as the \emph{discontinuity} region. The electric field on this plane, denoted as $\mathbf{E}_{\text{s}}(x,y,t)$, plays a central role in the model derivation and will be discussed in Sect.~\ref{sec:Sect2_2}.

The internal periodicity of the structure allows the electromagnetic fields in regions (1) and (2) to be expressed through a Floquet expansion of spatiotemporal harmonics. By means of Floquet’s theorem \cite{Floquet1883}, the analysis can be restricted to a single unit cell, which significantly reduces the overall complexity. Following \cite{Berral2015}, and considering first the purely spatial case (i.e., without time modulation), the electric field associated with a Floquet harmonic can be written as (using the convention $\mathrm{e}^{\mathrm{j}\omega t}\mathrm{e}^{-\mathrm{j}\vec{\mathbf{k}}\cdot\vec{\mathbf{r}}}$)
\begin{equation}\label{E_spatial}
\mathbf{e}_{nm}^{\text{cp/xp}, (v)}(x, y) = \frac{\hat{\mathbf{u}}_{nm}^{\text{cp/xp}}}{\sqrt{p_{\text{x}}p_{\text{y}}}} \text{e}^{\text{j}\omega_{0} t}\,  \text{e}^{-\text{j}( k_{n}x + k_{m}y +  \beta_{nm}^{(v)}z) }\, .
\end{equation}
Here, the indices $n$ and $m$ correspond to the spatial periods $p_{\text{x}}$ and $p_{\text{y}}$ along the $\hat{\mathbf{x}}$ and $\hat{\mathbf{y}}$ directions, respectively. The index ($v$) denotes the input (1) and output (2) regions. The quantities $k_n$ and $k_m$ refer to the transverse wavenumbers of the $(n,m)$th harmonic, while $\beta_{nm}^{(v)}$ is its longitudinal propagation constant, 
\begin{equation}
\beta_{nm}^{(v)} = \sqrt{\varepsilon_{\text{r}}^{(v)} \mu_{\text{r}}^{(v)}\frac{\omega_{0}^2}{c^2} - k_{n}^2 - k_{m}^2}, \hspace{5mm} v = 1, 2.
\end{equation}
The polarization state is described by the unit vector $\hat{\mathbf{u}}_{nm}^{\text{cp/xp}}$, where the superscripts $\text{cp}$ and $\text{xp}$ indicate co- and cross-polarized components (associated with TE/TM polarizations). The co-polarized component is defined with respect to the polarization of the incident wave. Explicit expressions for these parameters are provided in the Supplementary Material.

The inclusion of time modulation leads to a generalization of the harmonic definition in eq.~\eqref{E_spatial}. Since the time modulation is periodic, a discrete set of frequencies $\omega_i$ will be excited \cite{Alex2022TV}:
\begin{equation}\label{harmonic_freq}
\omega_{i} = \omega_{0} + \frac{2 i \pi}{T_{\text{M}}}, \hspace{5mm} i \in \mathbb{Z},
\end{equation}
where $\omega_{0}$ is the frequency of the incident wave and $T_{\text{M}}$ denotes the temporal period governing the modulation (see \cite{Moreno2023} and the Supplementary Material for a detailed discussion on the definition of this period and the concept of macroperiod). Here, $i$ is an integer index representing a certain $i$th temporal harmonic and should not be confused with the imaginary unit, denoted as $\mathrm{j}$. 

Accordingly, the electric field $\mathbf{e}_{nmi}$ associated with the individual $(n,m,i)$ spatiotemporal harmonic can be written as
\begin{equation}\label{E_ST}
\mathbf{e}_{nmi}^{\text{cp/xp}, (v)}(x, y, t) = \frac{\hat{\mathbf{u}}_{nm}}{\sqrt{p_{\text{x}}p_{\text{y}}T_{\text{M}}}}\, \text{e}^{\text{j} \omega_{i} t} \, \text{e}^{-\text{j} (k_{n}x + k_{m}y +  \beta_{nmi}^{v} z)}.
\end{equation} 
where the general expression for the propagation constant reads
\begin{equation}
\beta_{nmi}^{(v)} = \sqrt{\varepsilon_{\text{r}}^{(v)}\mu_{\text{r}}^{(v)}\frac{\omega_{i}^{2}}{c^2} - k_{n}^{2} - k_{m}^{2}},   \hspace{5 mm} v = 1,2.
\end{equation}
A dual definition exists for the magnetic-field vector associated with the same harmonic order:
\begin{equation}
\mathbf{h}_{nmi}^{\text{cp/xp}}(x, y, t) = \mathbf{\hat{z}} \times \mathbf{e}_{nmi}^{\text{cp/xp}}\,.
\end{equation}

In the space–time periodic scenario, the electromagnetic field in regions (1) and (2) can be expressed as a Floquet–Bloch expansion of spatiotemporal harmonics. The electric field is given by 
\begin{equation}\label{Efield}
\mathbf{E}(x, y, z, t) = \mathbf{e}_{000}^{\text{cp}} + \displaystyle\sum_{n,m,i} E_{nmi}^{\text{cp}} \mathbf{e}_{nmi}^{\text{cp}} + \displaystyle\sum_{n,m,i} E_{nmi}^{\text{xp}} \mathbf{e}_{nmi}^{\text{xp}},
\end{equation}
where the summation extends from $-\infty$ to $\infty$ over all indices. The vector $\mathbf{e}_{00}^{\text{cp}}$ represents the field vector of the incident wave (unit amplitude is assumed), while the unknown coefficients $E_{nmi}^{\text{cp/xp}}$ denote the amplitudes of the remaining harmonics contributing to the total field, including both co- and cross-polarized components. Among them, the co-polarized coefficient of order $n=m=i=0$, $E_{000}^{\text{cp}} = R$, corresponds to the reflection coefficient in region (1). In region (2), the transmission coefficient is obtained from the superposition of the incident and transmitted zeroth-order harmonics, yielding $T = 1 + E_{000}^{\text{cp}}$. 

A similar expression is proposed for the magnetic field $\mathbf{H}(x, y, z, t)$ at the input and output regions:
\begin{equation}\label{Hfield}
\mathbf{H}(x, y, z, t) = \mathbf{h}_{000}^{\text{cp}} + \displaystyle\sum_{n,m,i} H_{nmi}^{\text{cp}} \mathbf{h}_{nmi}^{\text{cp}} + \displaystyle\sum_{n,m,i} H_{nmi}^{\text{xp}} \mathbf{H}_{nmi}^{\text{xp}},
\end{equation}
where the amplitude coefficients of each magnetic-field harmonic $H_{nmi}^{\text{cp/xp}}$ are related to those of the electric field expansion as
\begin{equation}
H_{nmi}^{\text{cp/xp}} = Y_{nmi}^{\text{cp/xp}} \cdot E_{nmi} ^{\text{cp/xp}},
\end{equation}
with $Y_{nmi}^{\text{cp/xp}}$ being the modal admittance coefficients of order $nmi$.

From the previous definitions, it is clear that the body of unknowns to be resolved is the amplitudes $E_{nmi}^{\text{xp}}$. This discrete set of field amplitudes is, apparently, infinite. However, in practice, larger orders barely contribute, so after certain convergence tests, finite extensions can be accurate enough. To find the proper values of the unknowns, two important boundary conditions must be imposed. First, the electric field must be continuous across the spatial discontinuity $z = 0$,    
\begin{equation}\label{Econt}
\mathbf{E}(x,y,z = 0, t) = \mathbf{E}_{\text{s}}(x, y, t)\,. 
\end{equation}
where $\mathbf{E}(x,y,z = 0, t)$ is either the field expansion in (1)/(2). The electric field $\mathbf{E}_{\text{s}}(x, y, t)$ is the space-time field profile at the metasurface plane. This space-time profile depends on the scatterer shape and size \cite{Berral2015}, as well as on the introduced time modulation \cite{Alex2022TV}. As previously mentioned, the nature of this field will be largely discussed in Sect.~\ref{sec:Sect2_2}.  

\begin{figure}[t!]
  \centering
    \subfigure[]{\includegraphics[width= 0.9\columnwidth]{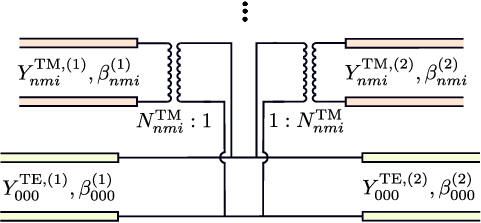}}
    \subfigure[]{\includegraphics[width= 0.9\columnwidth]{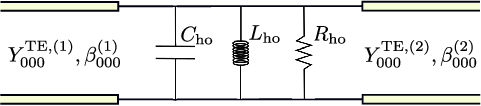}}
  \caption{(a): Equivalent circuit model capturing wave propagation in the space–time metasurface. TE incidence is assumed. The picture shows only a single higher-order harmonic of TM nature. The dots symbolize an infinite network of lines, encompassing both TE and TM modes. (b): Reduced version of the circuit, where the parallel connection of a capacitor, an inductor, and a resistor can emulate the presence of the higher-order harmonics.}
  \label{fig:circuit}
\end{figure}

Secondly, the continuity of the Poynting vector at the discontinuity is imposed as well,
\begin{multline}\label{Scont}
\mathbf{E}_{\text{s}}(x, y, t) \times [\mathbf{H}^{(1)}(x, y, z=0, t)] \\= \mathbf{E}_{\text{s}}(x, y, t) \times [\mathbf{H}^{(2)}(x, y, z=0, t)]\,.
\end{multline}
After introducing the field expansions \eqref{Efield} and \eqref{Hfield} in \eqref{Econt} and \eqref{Scont}, close-form expressions for the reflection coefficient $E_{000}^{\text{cp}} = R$ and the rest of harmonics are achieved, 
\begin{align}
\label{reflection} R &= \frac{Y_{000}^{\text{cp}} - Y_{000}^{\text{cp}} - Y_{\text{eq}} }{Y_{000}^{\text{cp}} + Y_{000}^{\text{cp}} + Y_{\text{eq}}} \\
\label{Enmi} E_{nmi}^{\text{cp/xp}} &= (1 + R) N_{nmi}^{\text{cp/xp}} \,.
\end{align}
From \eqref{reflection} the parameter $Y_{\text{eq}}$ is identified. It is internally defined in terms of the rest of the higher-order admittances as
\begin{equation}\label{Yeq}
Y_{\text{eq}} = \displaystyle\sum_{n, m, i}' |N_{nmi}|^{2}\, Y_{nmi}^{\text{cp}}  + \displaystyle\sum_{n, m, i} |N_{nmi}|^{2}\, Y_{nmi}^{\text{cp}}\,.
\end{equation}
where the symbol $\displaystyle\sum_{n,m,i}'$ indicates that the order $n = m = i = 0$ is excluded in the summation. Otherwise from \eqref{Enmi} and \eqref{Yeq} one finds the coefficient $N_{nmi}^{\text{cp/xp}}$, formally expressed as
\begin{equation}\label{transformador}
N_{nmi}^{\text{cp/xp}} = \frac{\displaystyle\int_{p_{\text{x}}} \displaystyle\int_{p_\text{y}} \displaystyle\int_{T_{\text{M}}} \mathbf{E}_{\text{s}}(x, y, t) \cdot \mathbf{e}_{nmi}^{\text{cp/xp}} \text{d}x \, \text{d}y \, \text{d}t }{\displaystyle\int_{p_{\text{x}}} \displaystyle\int_{p_\text{y}} \displaystyle\int_{T_{\text{M}}} \mathbf{E}_{\text{s}}(x, y, t) \cdot \mathbf{e}_{000}^{\text{cp}} \text{d}x \,\text{d}y \, \text{d}t}\,.
\end{equation}
From eq.~\eqref{Yeq}, the \emph{infinite} summation of admittances can be interpreted, from a circuit perspective, as an \emph{infinite} parallel connection of elements, each represented by an individual transmission line corresponding to the $(n,m,i)$th harmonic \cite{Moreno2024Eucap}. The parameter $N_{nmi}^{\text{cp/xp}}$ acts as a transformer that quantifies the coupling between the incident wave and the corresponding harmonic. This construction leads to the complete circuit representation shown in Fig.~\ref{fig:circuit}(a), with the reduced form depicted in Fig.~\ref{fig:circuit}(b). In this reduced model, the capacitance $C_{\text{ho}}$ accounts for the collective contribution of all TM evanescent harmonics, the inductance $L_{\text{ho}}$ represents the analogous contribution of TE evanescent harmonics, and the resistor $R_{\text{ho}}$ models the effect of all propagating harmonics, including both TE and TM components. Further details are provided in the Supplementary Material. 

%---------------------------------------------------------
\subsection{\label{sec:Sect2_2} Metasurface space-time field profile $\mathbf{E}_{\text{s}}(x,y,t)$}
%- Diode ON / Diode OFF \\
%- switching time too short \\
%- Factorized profile $E(x)E(t)$ \\
%- E metal = 0
As stated above, the field profile $\mathbf{E}_{\text{s}}(x,y,t)$ is a macro basis function that describes the evolution of the electric field at the space-time discontinuity ($z=0$), i.e., at the position where the space-time metasurface is placed. The developed method in Section II.A requires $\mathbf{E}_{\text{s}}(x,y,t)$ to be known and set \emph{a priori}. 

When canonical geometries and materials are considered, it is generally possible to find simple mathematical expressions that accurately describe, in a wide range of frequencies, the actual field profile  $E_s$ that is excited in the structure. This has been the case for time-invariant frequency selective surfaces (FSSs) and metasurfaces composed of periodic arrays of rectangular or circular insertions  \cite{Rengarajan2005, RodriguezUlibarri2017, CrosspolCircuit2021, MesaMosig2016}, and even for one-dimensional time-modulated diffraction gratings \cite{Alex2022TV, Moreno2023, Salva2024}.

When the considered geometries and materials are more intricate and simple mathematical expressions for $\mathbf{E}_{\text{s}}(x, y, t)$ cannot be formulated, it is still possible to rely on external numerical computations (e.g., CST, HFSS, or COMSOL) to extract the field profile. Just as with an analytical field profile, the numerical profile $\mathbf{E}_{\text{s}}(x, y, t)$ is inserted into \eqref{transformador} in Section II.A to compute the electromagnetic response of the system. This hybrid analytical-numerical technique has been previously employed to study time-invariant metasurfaces with complex geometries that act as filters and polarizers \cite{MesaMaria2018, MoleroMaria2019, Alex2021}. We will follow this approach when analytical field profiles cannot be obtained. 

Regardless of whether analytical or numerical field profiles are used, the space-time metasurface described in Fig.~\ref{fig:general} is mainly constituted by metallic insertions. In the extraction of $\mathbf{E}_{\text{s}}(x, y, t)$, we model the metallic sections of the metasurface as perfect electric conductors (PEC). In PEC regions, the tangential electric field nulls, thus $\mathbf{E}_{\text{s}}(x,y,t) = 0$. Moreover, PEC conditions imply instantaneous responses; namely, no transients apply. The assumption of PEC greatly simplifies the analysis of the space-time metasurface and, as discussed in \cite{ZapataTV2024}, is valid in a wide range of frequencies: from DC to the low-THz regime (up to the far infrared). At these frequencies, losses and transients in good conductors such as aluminum, copper, silver, or gold are negligible.

The last consideration regarding the nature of the field profile $\mathbf{E}_{\text{s}}(x,y,t)$ concerns its space-time dependence. Our previous studies on the matter showed that the spatial and temporal dependence in related time-varying metallic metasurfaces can be decoupled \cite{Salva2024}. Thus, spatial variations in the metasurface do not affect temporal ones and vice versa. From a mathematical perspective, this implies that the field profile can be generally factored as $\mathbf{E}_{\text{s}}(x,y,t) = \mathbf{E}_1(x,y)E_2(t)$, where $\mathbf{E}_1(x,y)$ and $E_2(t)$ are only functions of space and time, respectively. Having studied similar configurations and understanding the operation of the PIN diode, we believe the spatial and temporal dependence of our metasurface can be factored out. The hybrid configuration can still be applied. Since $\mathbf{E}_{1}(x, y)$ describes the spatial distribution of the field profile only, it can be analytically formulated if the field at the discontinuity is compatible with the entire-domain function, or otherwise it can be extracted from CST when complex scatterers are regarded. The field profile $\mathbf{E}_{1}(x, y)$ is that obtained in a steady-state scenario. The time evolution $E_{2}(t)$ is later incorporated once the time modulation is known.

%-----------------------------------
\subsubsection{Presence and Impact of the PIN diode}
\begin{figure}[t!]
  \centering
    \includegraphics[width= 0.8\columnwidth]{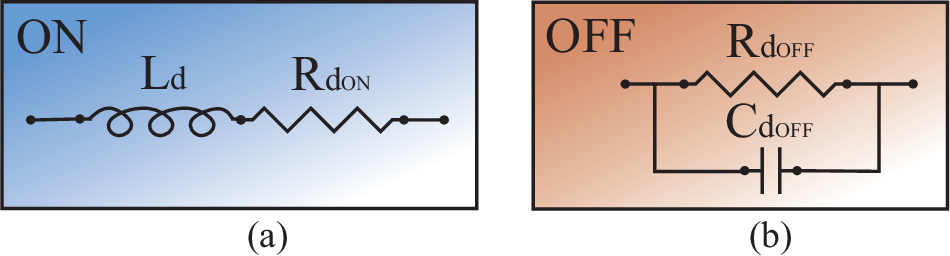}
  \caption{Equivalent circuits for the PIN diode: (a) ON state, (b) OFF state.}
  \label{fig:equivalentdiode}
\end{figure}

Regarding the PIN diodes and their impact on the field profile, they can be modeled at RF and microwave frequencies using the simplified equivalent circuit shown in Fig.~\ref{fig:equivalentdiode} \cite{MarioDiodes2024}. A PIN diode essentially behaves as a switch with two operating states: ON and OFF. In the ON state, the diode is modeled as a series connection of an inductor $L_\mathrm{d}$ and a resistor $R_\mathrm{dON}$. In the OFF state, it is represented by a parallel combination of a resistor $R_\mathrm{dOFF}$ and a capacitor $C_\mathrm{dOFF}$.

In this work, the commercial PIN diode MACOM MA4AGP907~\cite{MACOMMA4AGP907} is considered. According to values reported in the literature~\cite{WanDiodo2021,AngelDiodo2025}, obtained through characterization procedures, the equivalent circuit parameters are $L_\mathrm{d}=50$~pH, $R_\mathrm{dON}=4.2$~$\Omega$, $C_\mathrm{dOFF}=42$~fF, and $R_\mathrm{dOFF}=300$~k$\Omega$. Using these parameters, the equivalent impedances of the diode at 1 and 10 GHz are listed in Table~\ref{tab:diodestates}. As observed, the impedance in the ON state is significantly lower, indicating that the diode behaves similarly to a lossy metal. In contrast, the impedance in the OFF state is much higher, making the diode approximately equivalent to an open circuit. However, this approximation is valid only at relatively low frequencies, since the impedance in the OFF state decreases significantly as frequency increases.

Another important factor to consider is the diode switching speed. According to the manufacturer's datasheet, the switching time measured between 10\% and 90\% of the RF voltage, or between 90\% and 10\%, for a single series-mounted diode is approximately 2~ns (measured at 10~GHz). Consequently, modulation signals with periods shorter than about 4~ns cannot be reliably implemented, corresponding to a maximum modulation frequency of roughly 250~MHz. Given the limitations of commercial PIN diodes, their use for validating the proposed model is therefore restricted to relatively low frequencies, typically on the order of several megahertz or a few gigahertz, where the diode can be reasonably approximated as a perfect electric conductor in the ON state and as an open circuit (air) in the OFF state. In addition, switching times shorter than a few nanoseconds cannot be achieved with this technology; achieving faster transitions would require alternative devices, such as photodiodes~\cite{JonesTimeReflection2024}. Nevertheless, despite these limitations, this approach provides a practical means of experimentally validating the model, since it enables the implementation of a circuit with sufficiently fast switching while maintaining reasonably accurate ON--OFF state approximations during the fabrication and measurement of a surface based on the studied topology.

\begin{table}[h!]
\caption{Impedance of MA4AGP907 PIN diode at different frequencies.}
\label{tab:diodestates}
\begin{tabular}{ccc}
\hline
State & @1 GHz & @10 GHz \\ \hline
ON    &  4.21$e^{j4.27^o}$    &  5.24$e^{j36.79^o}$       \\ \hline
OFF   &  3789.1$e^{-j89.27^o}$     &  378.94$e^{-j89.92^o}$      \\ \hline
\end{tabular}
\end{table}

%------------------------------------
\section{\label{sec:subSect3_1} Numerical results}

Now that the operational domain and inherent limitations of the frequency-domain framework have been clearly established, we employ it to investigate a set of two-dimensional time-modulated metasurface configurations incorporating PIN diodes, to uncover physical phenomena of potential relevance for engineering applications. In particular, this approach enables the systematic exploration of spatiotemporal scattering effects, frequency conversion mechanisms, and nonreciprocal responses arising from the imposed modulation schemes. Although the proposed framework is primarily tailored for (2$+$1)-D scenarios, for the sake of completeness and to facilitate validation, the (1$+$1)-D cases previously developed by the authors are also included in the Supplementary Material, where they serve as benchmark examples supporting the accuracy and consistency of the model.

%-----------------------------------------------------------
\subsection{\label{sec:subSect3_2}Case 1: Time-modulated Inductive Mesh}

\begin{figure}[!t]
    \centering    \includegraphics[width=1\columnwidth]{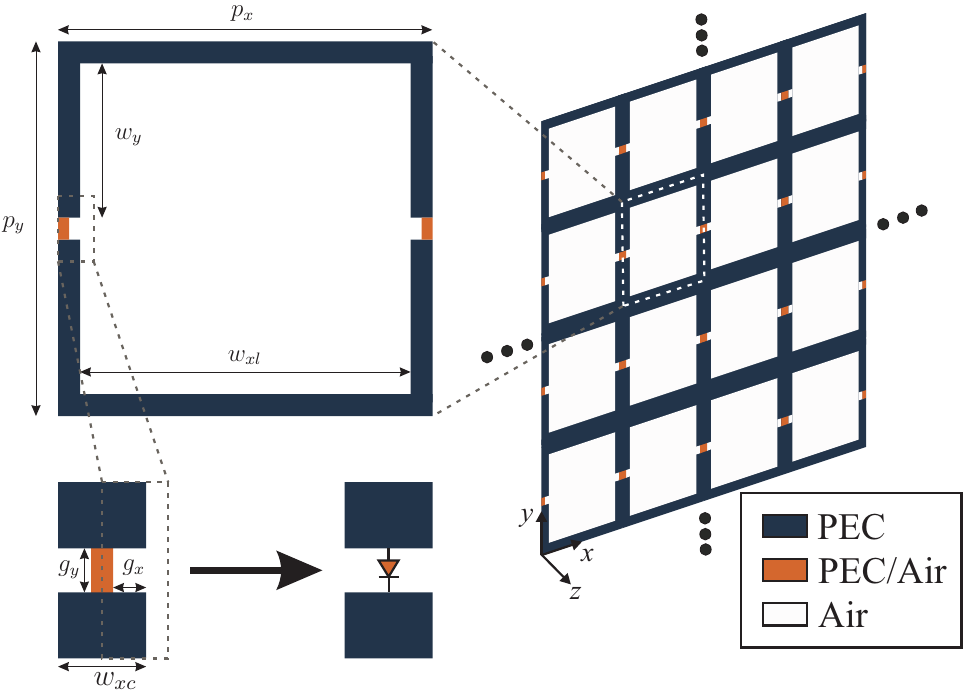}
    \caption{Unit cell geometry including the integrated PIN diode. The metasurface is periodically repeated with $p_{\text{x}} = p_{\text{y}} = 20\,\mathrm{mm}$. The central aperture dimensions are $w_{\text{xl}} = 17\,\mathrm{mm}$ and $w_{\text{y}} = 8.35\,\mathrm{mm}$, while the central slot width is $w_{\text{xc}} = 3\,\mathrm{mm}$. The diode dimensions are $g_{\text{y}} = 0.67\,\mathrm{mm}$ and $g_{\text{x}} = 1.32\,\mathrm{mm}$, according to the physical dimensions of the employed PIN diode (MA4AGP907).}
    \label{estructura}
\end{figure}

The first space-time metasurface under analysis is depicted in Fig.~\ref{estructura}. The core structure consists of an electrically thin metallic sheet (zero thickness from a theoretical perspective), perforated by a periodic arrangement of rectangular apertures. This part can essentially be interpreted as an inductive-mesh frequency-selective surface (FSS) \cite{costa2014overview}. Then, two narrow gaps are introduced along the lateral regions of each unit cell, creating metal-free zones that enable the integration of PIN diodes. A detailed view of this region is provided in the Fig.~\ref{estructura}, where the PIN diode is represented as a metallic inclusion with dimensions $g_{\text{y}} \times (w_{\text{xc}} - 2w_{\text{gx}})$, consistent with the adopted MA4AGP907 component. By appropriately biasing these elements, the effective surface impedance becomes time-dependent, thereby realizing a time-modulated inductive FSS. The metasurface is illuminated under normal incidence by a TE-polarized plane wave, with the electric field oriented along the $\hat{\mathbf{y}}$ direction, and, due to the geometrical symmetry of the unit cell with respect to the principal planes, no cross-polarized scattering is expected, so the response remains confined to the co-polarized component.

\begin{figure}[t!]
    \centering
    \subfigure[]{\includegraphics[width=0.49\columnwidth]{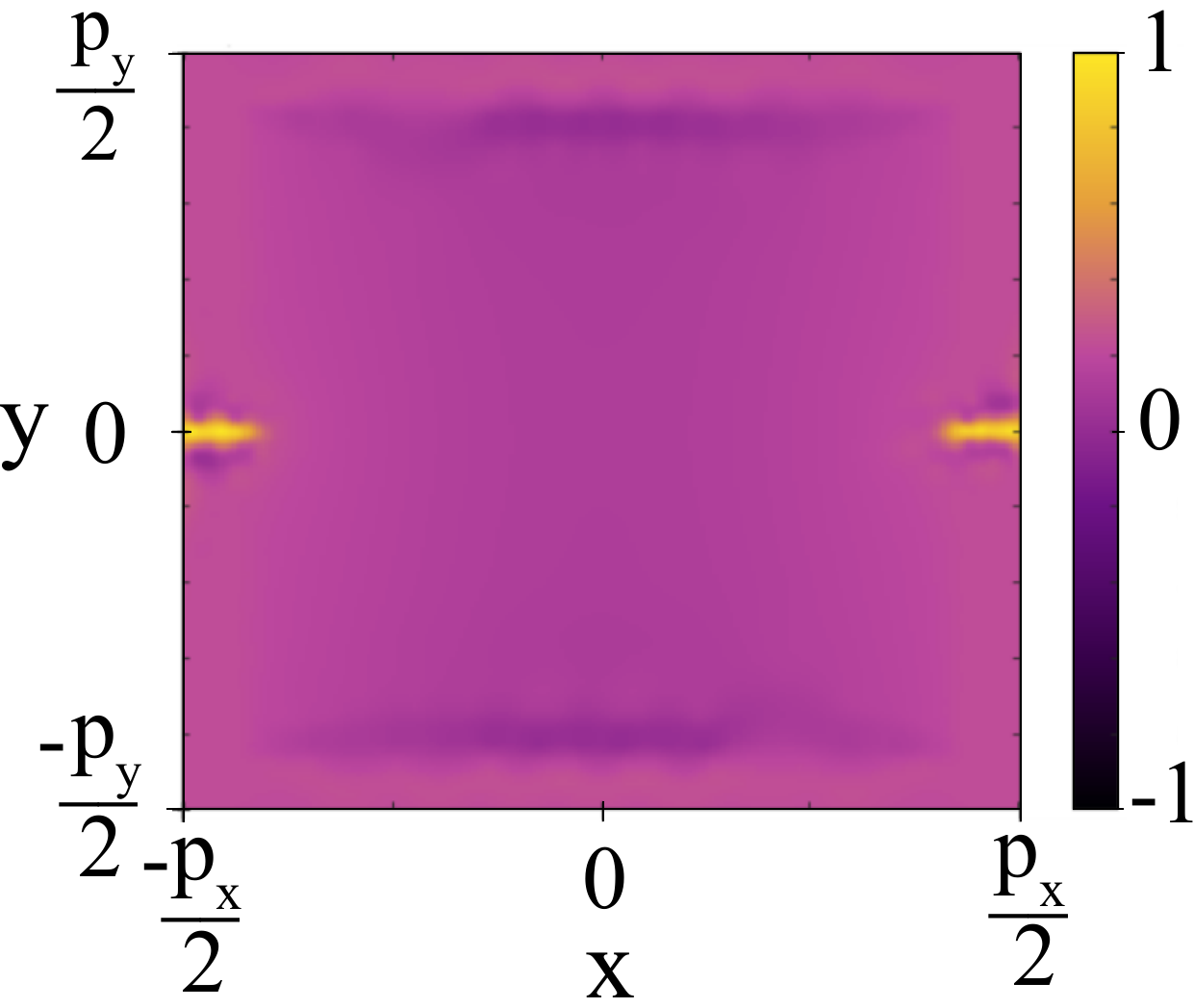}}
    \subfigure[]{\includegraphics[width=0.49\columnwidth]{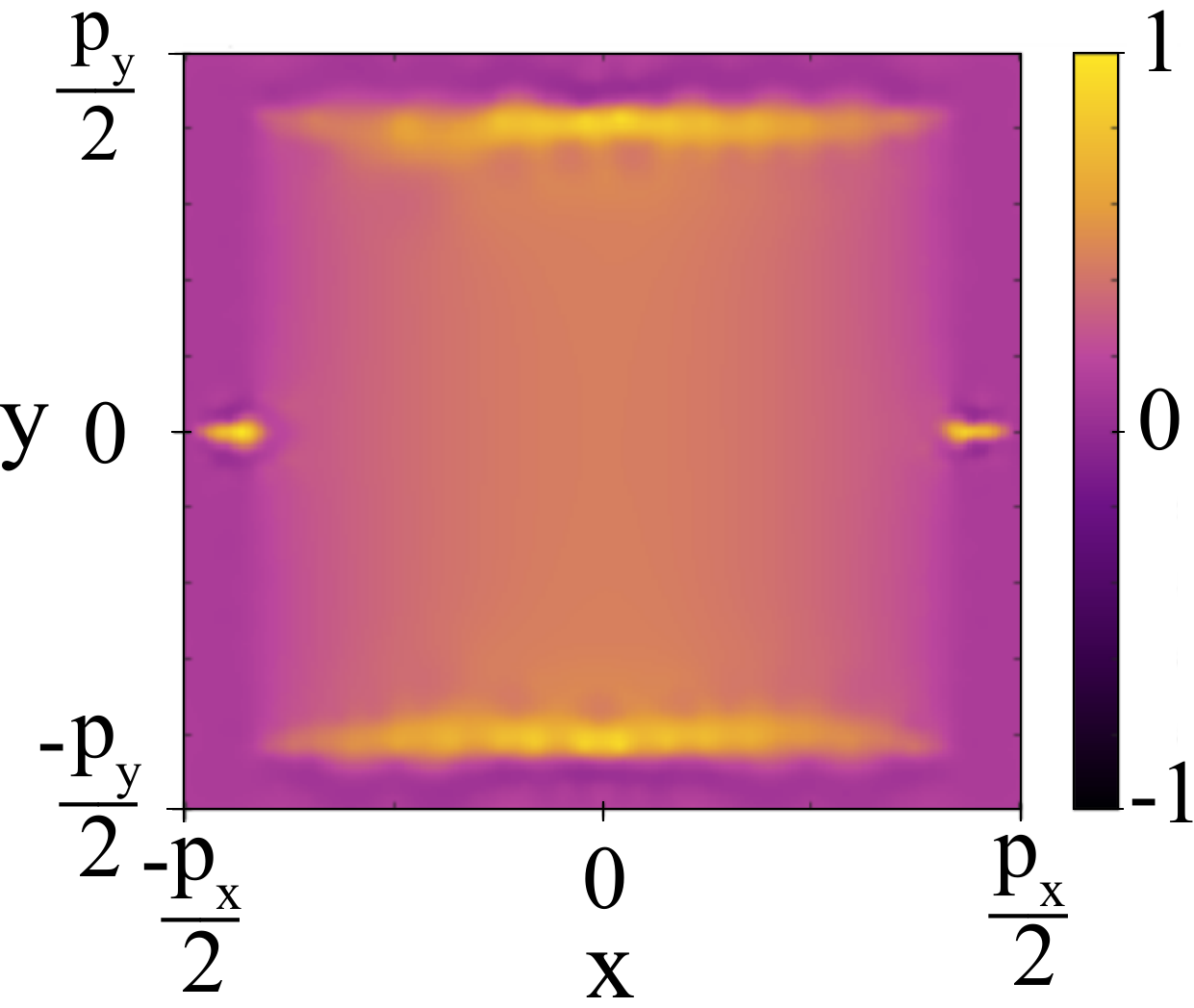}} \\ % Salto de línea para nueva fila
    \caption{Electric field profiles $\mathbf{E}_{\text{OFF/ON}}(x, y)$ in the (a) OFF, and (b) ON states, both extracted from CST. Parameters of the structure: $p_{\text{x}} = p_{\text{y}} = 20\,$mm, $w_{\text{xl}} = 17\,$mm, $w_{\text{xc}} = 3\,$mm, $w_{\text{y}} = 8.35\,$mm, $g_{\text{y}} = 0.67\,$mm, $g_{\text{x}} = 1.32\,$mm.}
    \label{fig:3D_original}
\end{figure}

Time modulation is achieved by periodically switching the diodes' states. In the time-invariant limit, where the diodes remain fixed in either the ON or OFF state, the metasurface exhibits a stationary response under plane-wave illumination. In contrast, a dynamic regime emerges when the diode state varies periodically. Assuming a modulation period $T_{\text{M}}$, the diode remains in the OFF state for $0 \le t < D T_{\text{M}}$ and in the ON state for $D T_{\text{M}} \le t < T_{\text{M}}$, where $0 \le D \le 1$ defines the \emph{duty cycle}. This parameter enables controlled breaking of temporal symmetry between the ON and OFF time intervals \cite{Moreno2023}, thereby introducing additional degrees of freedom that can be exploited to tailor the metasurface response.

The space-time metasurface field profile  $\mathbf{E}_{\text{s}}(x, y, t)$ takes the approximation described in Supplementary Material, with the spatial and temporal contributions being decoupled, 
\begin{equation}\label{scatt1_field}
\mathbf{E}_{\text{s}}(x, y, t) =  \sin(\omega_{0} t) \left\{
\begin{array}{ll}
\mathbf{E}_{\text{OFF}}(x, y) \hspace{5mm} t \in [0, DT_{\text{M}})   \\
\mathbf{E}_{\text{ON}}(x, y) \hspace{5mm} t \in [DT_{\text{M}}, T_{\text{M}})\, .  
\end{array}
\right.
\end{equation}

The temporal evolution is analytically described by the term $\sin(\omega_0 t)$, which provides a convenient and tractable representation of the modulation process. The spatial field distributions $\mathbf{E}_{\text{OFF/ON}}(x, y)$ are those represented in Fig.~\ref{fig:3D_original}. As discussed in the Supplementary Material, these spatial profiles can be obtained in closed form only for relatively simple scatterer geometries, where accurate analytical models of the electromagnetic response are available and provide valuable physical insight. For more intricate configurations, however, deriving such expressions can become cumbersome or even intractable due to the geometry's complexity and associated boundary conditions. In these cases, it is advantageous to complement the analytical framework with numerically extracted field distributions, for instance, obtained from full-wave simulations using CST. This hybrid approach, combining analytical temporal modeling with numerically derived spatial profiles, enables an accurate and realistic characterization of the metasurface response while preserving the interpretability of the formulation.

These profiles are those manifested in the static case, where the diode is permanently in the OFF or ON state. The effect of the diode is visible in the left- and right-most part of the cell. When the diode is in the OFF state, almost all the field is confined in the gap region, where the metals are closer. This field distribution is reorganized once the diode is switched ON. The closest metallic parts are now connected, and the field resembles that of a $\text{TE}_{10}$ in a rectangular waveguide. Both spatial profiles are combined during a time cycle, as expressed in \eqref{scatt1_field}.

\begin{figure}[t!]
    \centering
    \subfigure[\hspace*{-1.2cm}]{
    \includegraphics[width=\columnwidth]{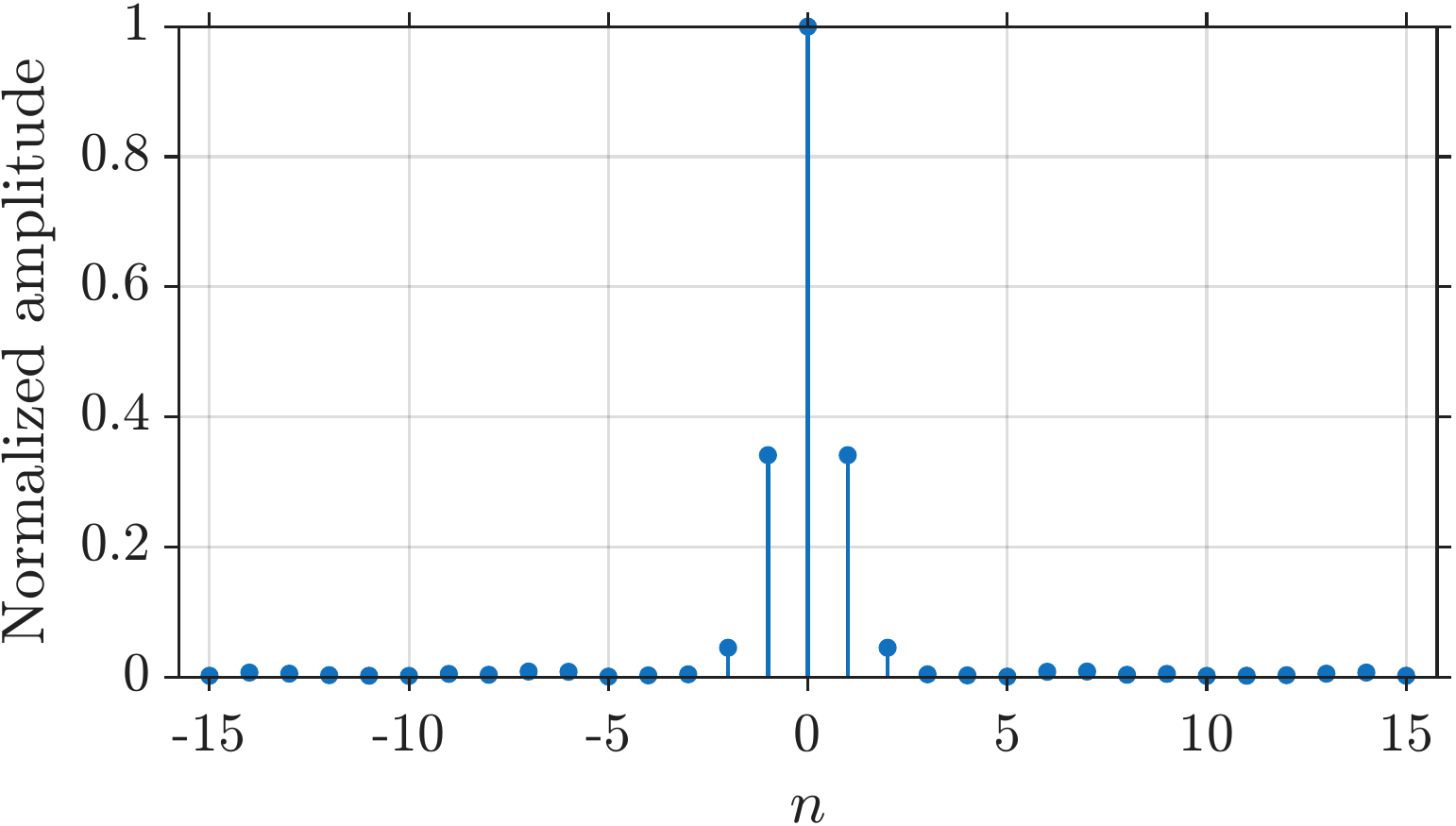}}
    \vspace{0.1cm}
    \subfigure[\hspace*{-1.2cm}]{
      \includegraphics[width=\columnwidth]{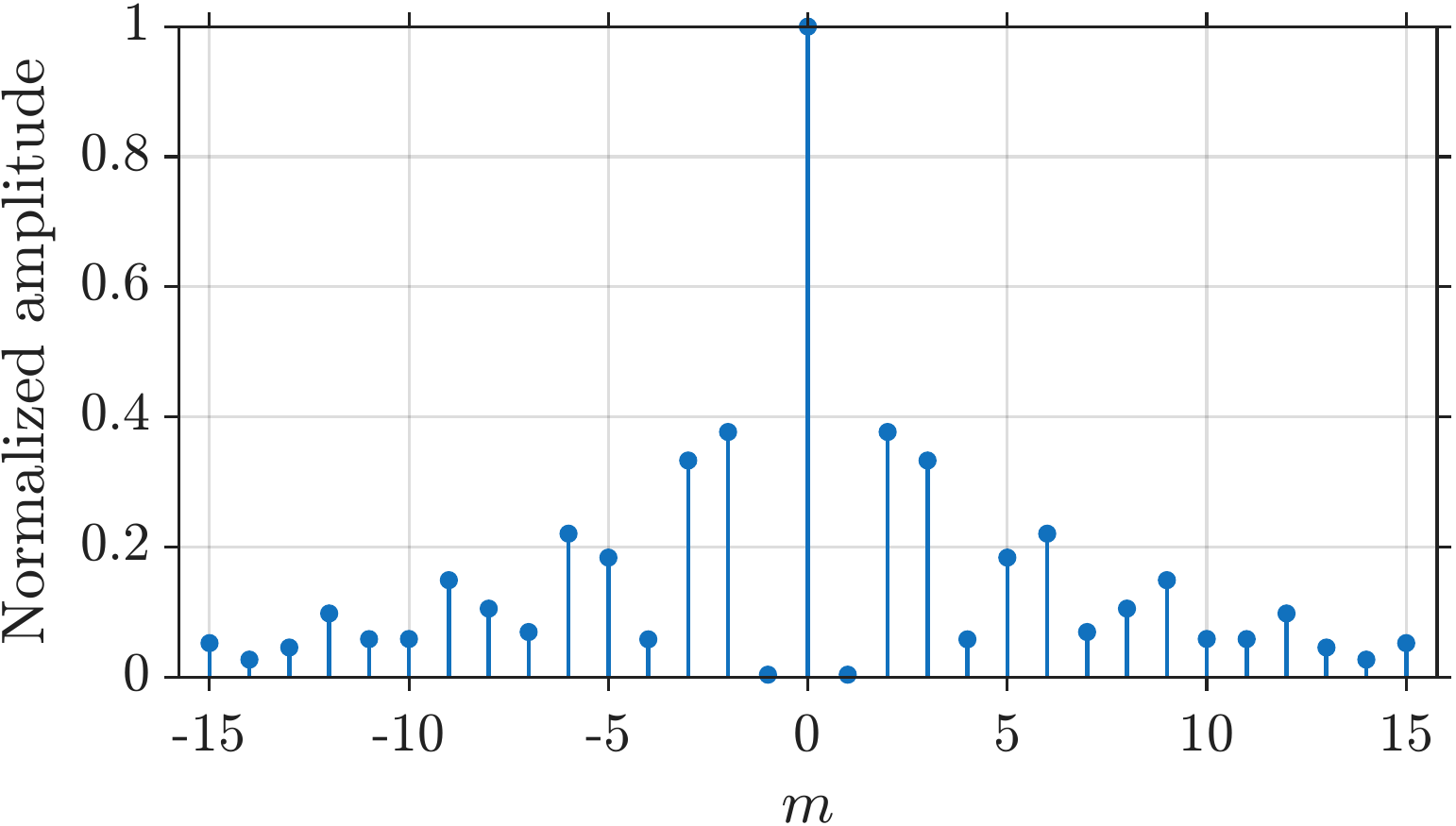}}
    \vspace{0.1cm}
    \subfigure[\hspace*{-1.2cm}]{
      \includegraphics[width=\columnwidth]{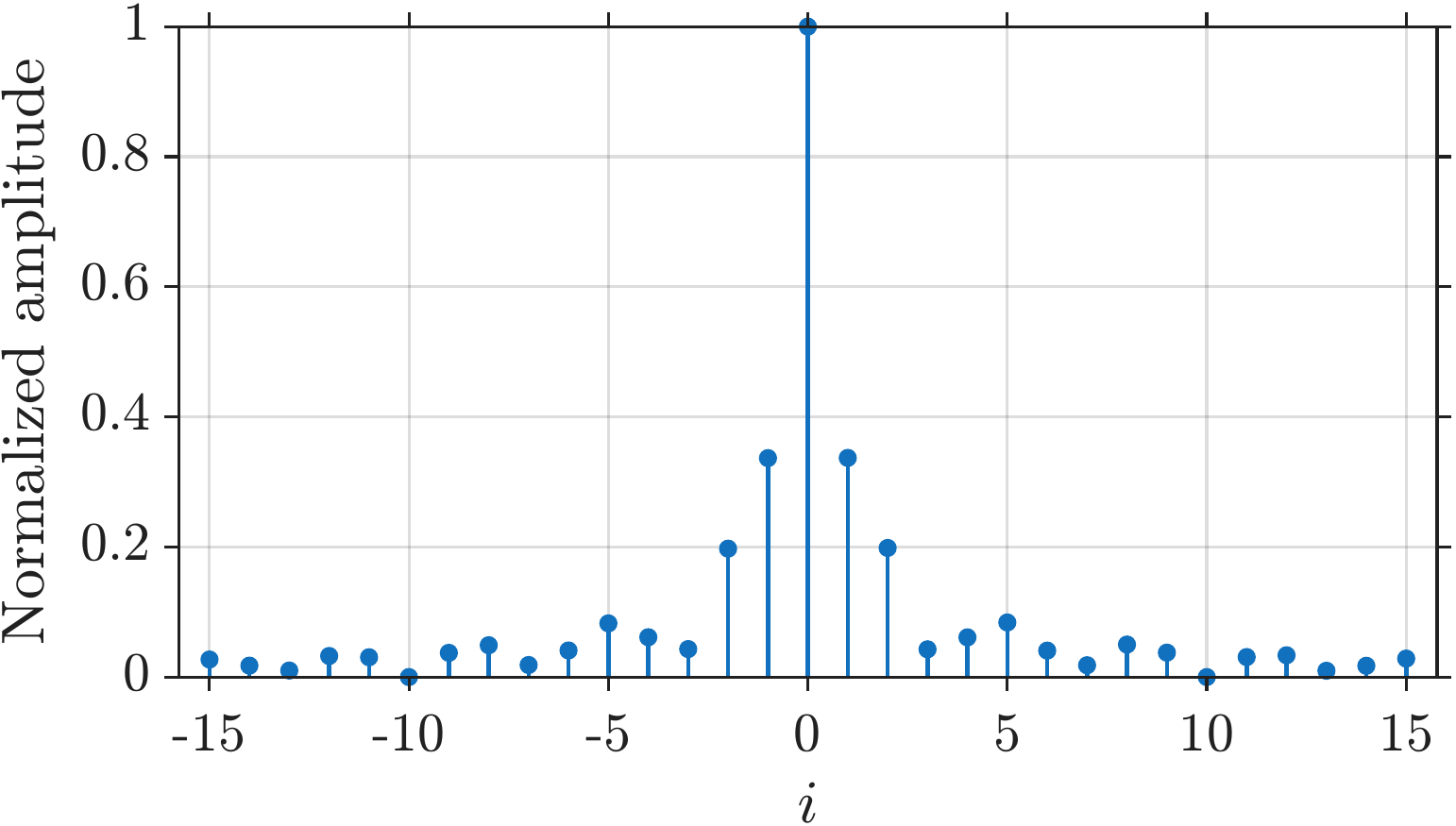}}
    \caption{Normalized amplitudes of representative space--time Floquet harmonics excited in the time-modulated inductive metasurface shown in Fig.~\ref{estructura}. (a) Spatial harmonics associated with the periodicity along the $x$-direction ($m=i=0$, $n\neq0$). (b) Spatial harmonics associated with the periodicity along the $y$-direction ($n=i=0$, $m\neq0$). (c) Temporal harmonics generated by the periodic modulation ($n=m=0$, $i\neq0$). The results show that most of the scattered power is concentrated around the fundamental harmonic, while higher-order spatial and temporal harmonics are also excited due to the imposed space--time modulation. Structure parameters are the same as in Fig.~\ref{estructura} and Fig.~\ref{fig:3D_original}. The metasurface is illuminated under normal incidence at $f_{0}=15\,\mathrm{GHz}$, with modulation period $T_\mathrm{M}=20\,\mathrm{ns}$ and duty cycle $D=0.3$.}
    
    \label{armonicos}
\end{figure}

\begin{figure}[t!]
    \centering
    \subfigure[]{\includegraphics[width=0.95\columnwidth]{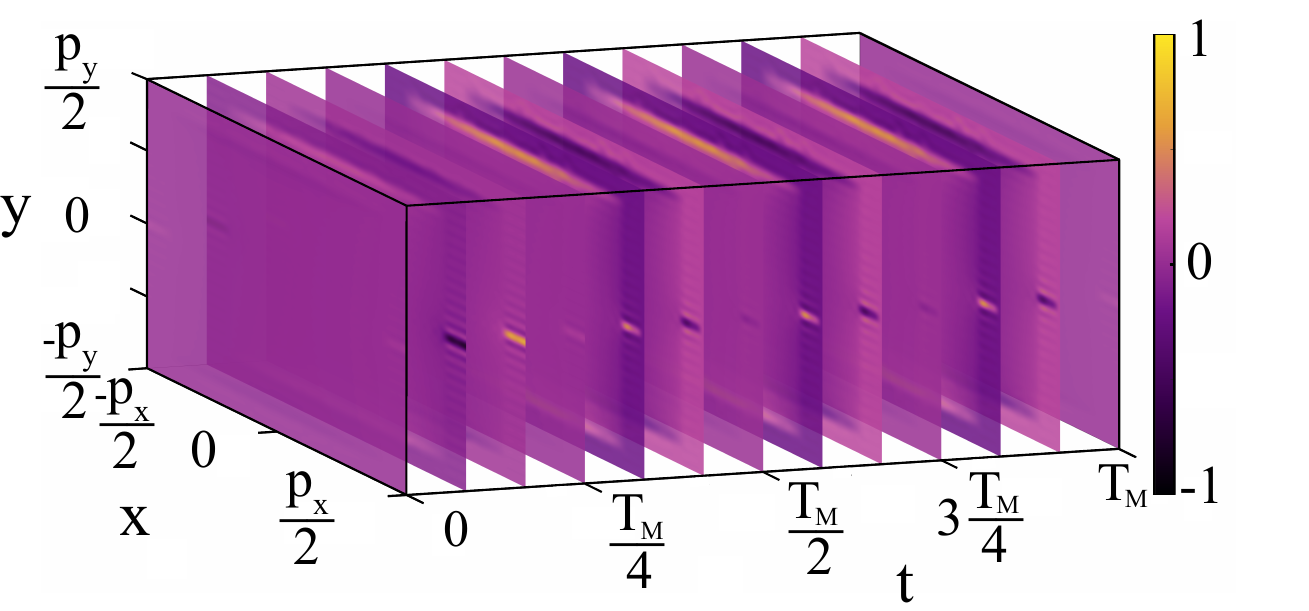}}
    \subfigure[]{\includegraphics[width=0.49\columnwidth]{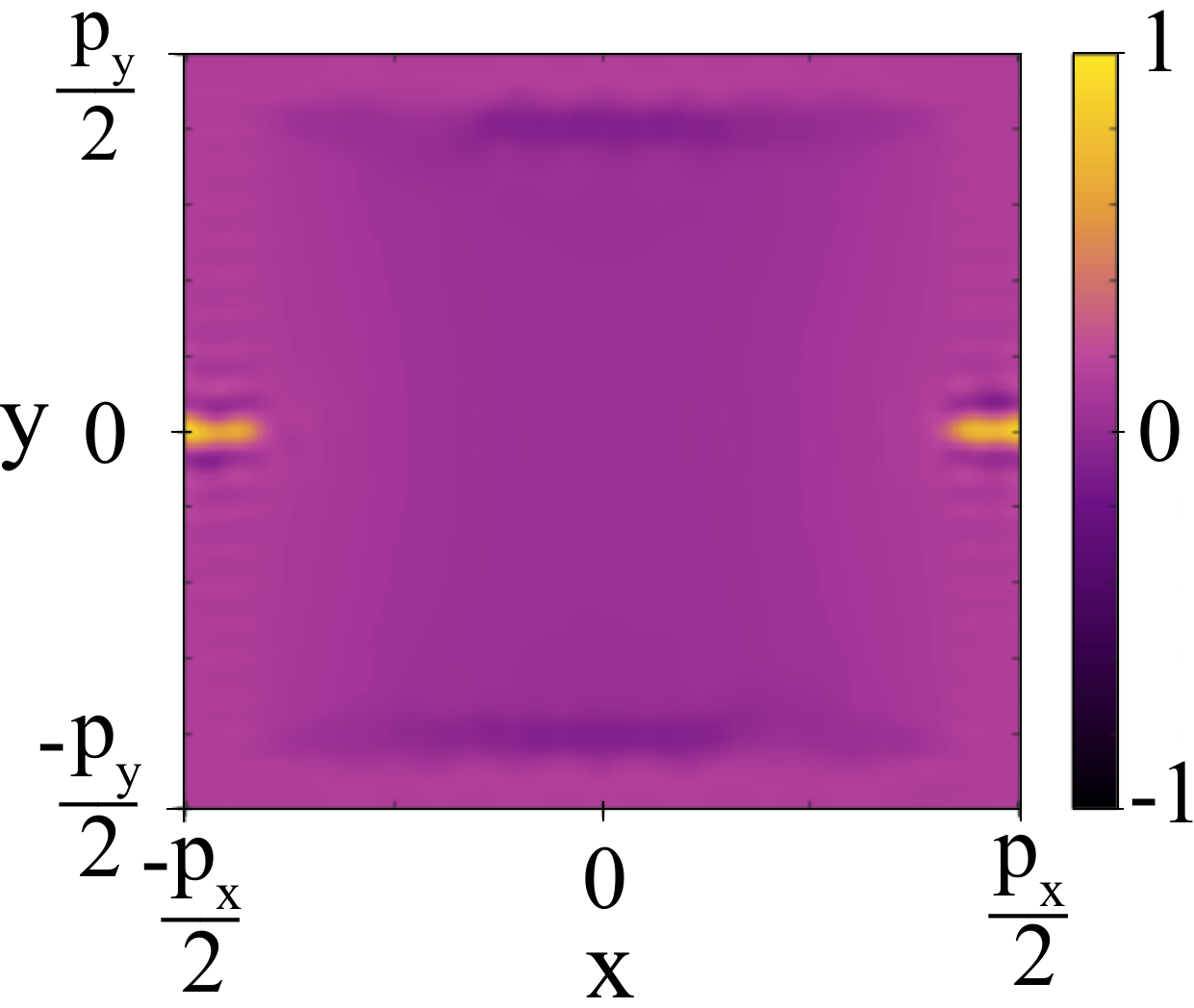}}
    \subfigure[]{\includegraphics[width=0.49\columnwidth]{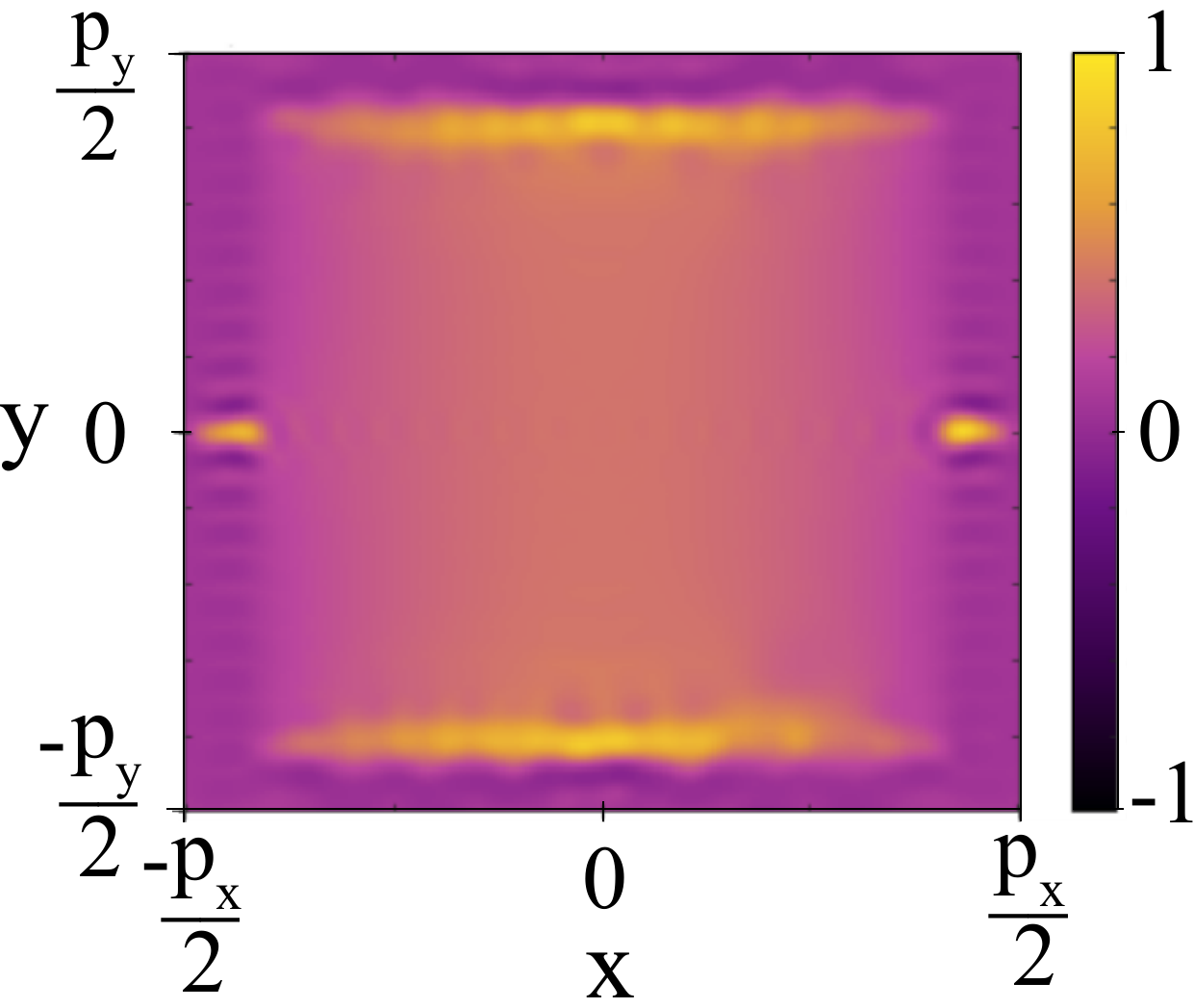}}
 \\ % Salto de línea para nueva fila
    \caption{(a) Reconstruction of the electric field $E^{(1)}(x, y, z = 0, t)$ at the discontinuity along a full temporal cycle $T_{\text{M}}$; (b) Reconstructed field $E^{(1)}(x, y, z = 0, t = T_{\text{M}}/8)$ at $t = T_{\text{M}}/8$; (c) Reconstructed field $E^{(1)}(x, y, z = 0, t = 3T_{\text{M}}/4)$ at $t = 3T_{\text{M}}/4$. The parameters of the Structure are the same as in Fig.~\ref{estructura}. The temporal period is now $T_{M} = 3T_{0}$. 
    }
    \label{fig:3D}
\end{figure}

Typical responses of space–time periodic structures are based on electromagnetic fields that are rich in harmonics. This behavior contrasts with stationary situations, where, in most cases, only the fundamental harmonic propagates. The space-time spectrum is formed by individual contributions of harmonics moving in different directions and vibrating with different frequencies. A visual representation of the amplitude associated with some harmonics is shown in Fig.~\ref{armonicos}. The structure parameters are included in the figure caption. For the evaluation of $E_{nmi}$, the reflection coefficient is previously evaluated by using \eqref{reflection} and introduced in \eqref{Enmi}. The structure is illuminated by a normally impinging plane wave with angular frequency $\omega_{0} = 2\pi \cdot 15\cdot 10^{9}\, \text{rad/s}$. The modulation periodicity is set at $T_{\text{M}} = 10\,$ns and the duty cycle is fixed at $D = 0.3$. 

Fig.~\ref{armonicos}(a) shows the amplitudes of the harmonics indexed by $n$, with $m = i = 0$ held fixed. This plot represents the set of spatial harmonics associated with the periodicity along $\hat{\mathbf{x}}$, where $k_{m=0} = 0$ and $\omega_{i=0} = \omega_0$, so frequency conversion is not considered. The dominant contribution corresponds to the $n = 0$ harmonic, indicating that most of the power is carried by the fundamental component. A similar scenario is illustrated in Fig.~\ref{armonicos}(b), where the spatial harmonics associated with the periodicity along $\hat{\mathbf{y}}$ are examined, with $m$ varying and $n = i = 0$ fixed. As in the previous case, the fundamental harmonic ($m = 0$) carries the largest portion of the power, while higher-order spatial harmonics contribute less significantly.

A more interesting case is that plotted in Fig.~\ref{armonicos}(c), where the indices $n = m = 0$ are fixed. This figure highlights the system's multi-frequency behavior. Although the fundamental harmonic still carries a significant portion of the power, newly generated temporal harmonics ($i \ne 0$) also receive part of the input power. The frequency of each of these harmonics is computed according to \eqref{harmonic_freq}, contributing to a global and multi-frequency scattered field. The proposed model accurately reproduces and predicts this behavior, representing a clear advancement over existing models in the literature.  

The model can be used to reconstruct the field profile at the discontinuity using the Floquet expansion in \eqref{Efield}. This is a consequence of the condition imposed in \eqref{Econt}. Fig.~\ref{fig:3D}(a) shows the field reconstruction frame by frame. Each frame denotes the field at a different time instant, inside the time interval $0 \le t \le T_{\text{M}}$. To avoid a large number of frames, the time periodicity has been increased up to $T_{\text{M}} = 3T_{0}$, and the duty cycle is set to be $D = 0.3$. This means that for $t < DT_{\text{M}}$, the field profile resembles that corresponding to the diode in OFF. This can be better visualized in Fig.~\ref{fig:3D}(b), where the field is evaluated in a frame at $t = T_{\text{M}}/8$. As expected, according to the stationary case in Fig.~\ref{fig:3D_original}(a), the maximum value is confined in the gap regions. For $t > DT_{\text{M}}$, the diode is in ON, and the field must be similar to the profile in Fig.~\ref{fig:3D_original}(b). Fig.~\ref{fig:3D}(c) evaluates the field at $t = 3T_{\text{M}}/4$, corroborating the expected predictions. It is worth remarking that the field in $\mathbf{E}(x, y, z = 0, t)$ in \eqref{Econt} is described by the infinite harmonic expansion in \eqref{Efield}. The triple infinite series is approximated by truncating the indices at $n, m, i \le 15$. The truncation effects become more noticeable at the lateral regions of the fields in Fig.~\ref{fig:3D}(b)–(c), where slight undulations are observed. Although this represents a slight deviation with respect to the \emph{exact} field profiles in Fig.~\ref{fig:3D_original}(a)-(b), the agreement is quite good.     

\subsection{Case 2: Time-modulated Slot Insertion}

A second application of the model focuses on the analysis of space--time properties in the vicinity of resonant behavior. In contrast to conventional one-dimensional, time-modulated structures analyzed in \cite{Salva2024, Moreno2025}, two-dimensional slotted apertures exhibit resonant responses characterized by full transmission of the incident wave at the resonance frequency. In this context, the temporal modulation acts as an additional degree of freedom, enabling an \emph{innovative mechanism} for tuning the transmission band.

\begin{figure}[t!]
    \centering
    \includegraphics[width=0.99\columnwidth]{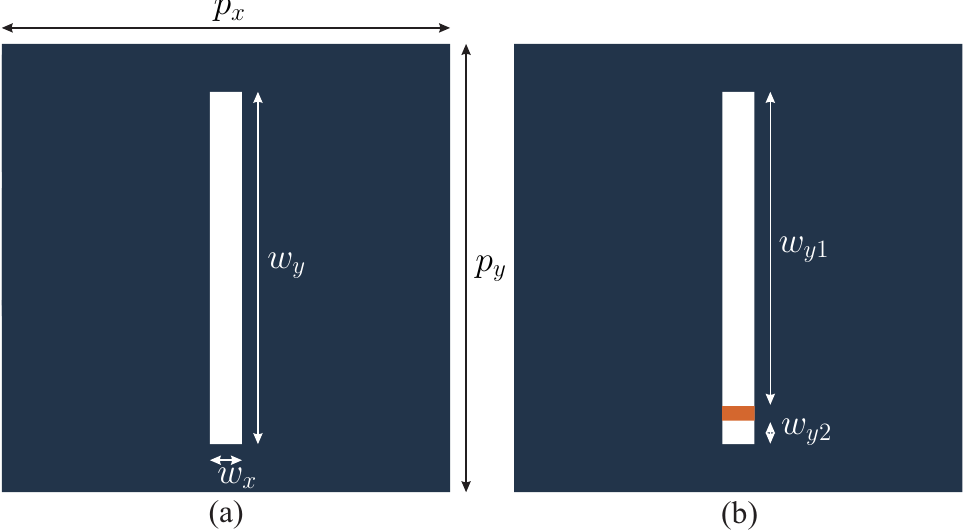}
    \caption{Unit cell of the time-modulated slot-insertion metasurface. The metasurface is periodically repeated with $p_x = p_y = 10\,\mathrm{mm}$. (a) OFF state, where the rectangular slot ($w_y = 8\,\mathrm{mm}$, $w_x = 0.3\,\mathrm{mm}$) is continuous. (b) ON state, where the PIN diode ($0.1\,\mathrm{mm}$ along $y$ and $0.3\,\mathrm{mm}$ along $x$) electrically divides the slot into two coupled apertures ($w_{y1} =7.375\,\mathrm{mm}$, $w_{y2} =0.325\,\mathrm{mm}$).}
    \label{ranura_dividida}
\end{figure}

To evaluate the ability of the proposed model to describe resonant behaviors in time-modulated structures, we consider a metasurface whose unit cell is shown in Fig. \ref{ranura_dividida}. The space-time metasurface is a holey metallic plate formed by the periodic repetition of rectangular apertures etched into the surface. The spatial periods are $p_x = p_y = 10\,\mathrm{mm}$. The thickness of the metasurface is negligible. The aperture dimensions are $w_y = 8\,\mathrm{mm}$ and $w_x = 0.3\,\mathrm{mm}$. Time modulation is induced by means of a PIN diode, placed at a certain height of the slot, $-w_{\text{y}}/2 + w_{\text{y}2}$, and with the same dimensions as the previous case.

The structure is fully symmetric with respect to the principal planes. A plane wave with the electric field directed along the $x$-direction illuminates the structure. 
When the diode is in the OFF state, it approximately behaves as an open circuit. Thus, the effective length of the rectangular slot is directly $w_y$, inducing a resonance when the wavelength is about $\lambda = 2 w_{\text{y}}$. In contrast, when the diode is in the opposite state, it behaves effectively as a metallic element, thus splitting the original single slot into two electromagnetically-coupled apertures. This substantially modifies the field distribution at the discontinuity plane and, consequently, the effective resonance frequency of the unit cell. In a steady-state scenario (with no time modulation), the full-transmission peak shifts in frequency with the diode's state. 

The dynamic scenario includes time modulation on the unit cell. It is implemented by periodically switching between the ON and OFF states along time. Considering a full cycle with periodicity $T_{\text{M}}$, $DT_{\text{TM}}$ indicates the time interval during which the diode remains in the OFF state. Accordingly, $[1-D]T_{\text{M}}$  corresponds to the time spent in the ON state along a temporal cycle. The electric-field profile at the discontinuity can be expressed again with \eqref{scatt1_field},
where the spatial profiles are extracted from CST in the steady-state. Please note that  $\mathbf{E}_{\text{OFF}}(x, y)$ and $\mathbf{E}_{\text{ON}}(x, y)$ take different values than in Sec. III.A, although the spatiotemporal field profile $\mathbf{E}_{\text{s}}$ is constructed similarly. The space-time profile in \eqref{scatt1_field} is then employed to estimate the transformers in \eqref{transformador}, enabling the use of the equivalent circuit in \eqref{fig:circuit}. 

\begin{figure}[t!]
    \subfigure{\centering\includegraphics[width=\columnwidth]{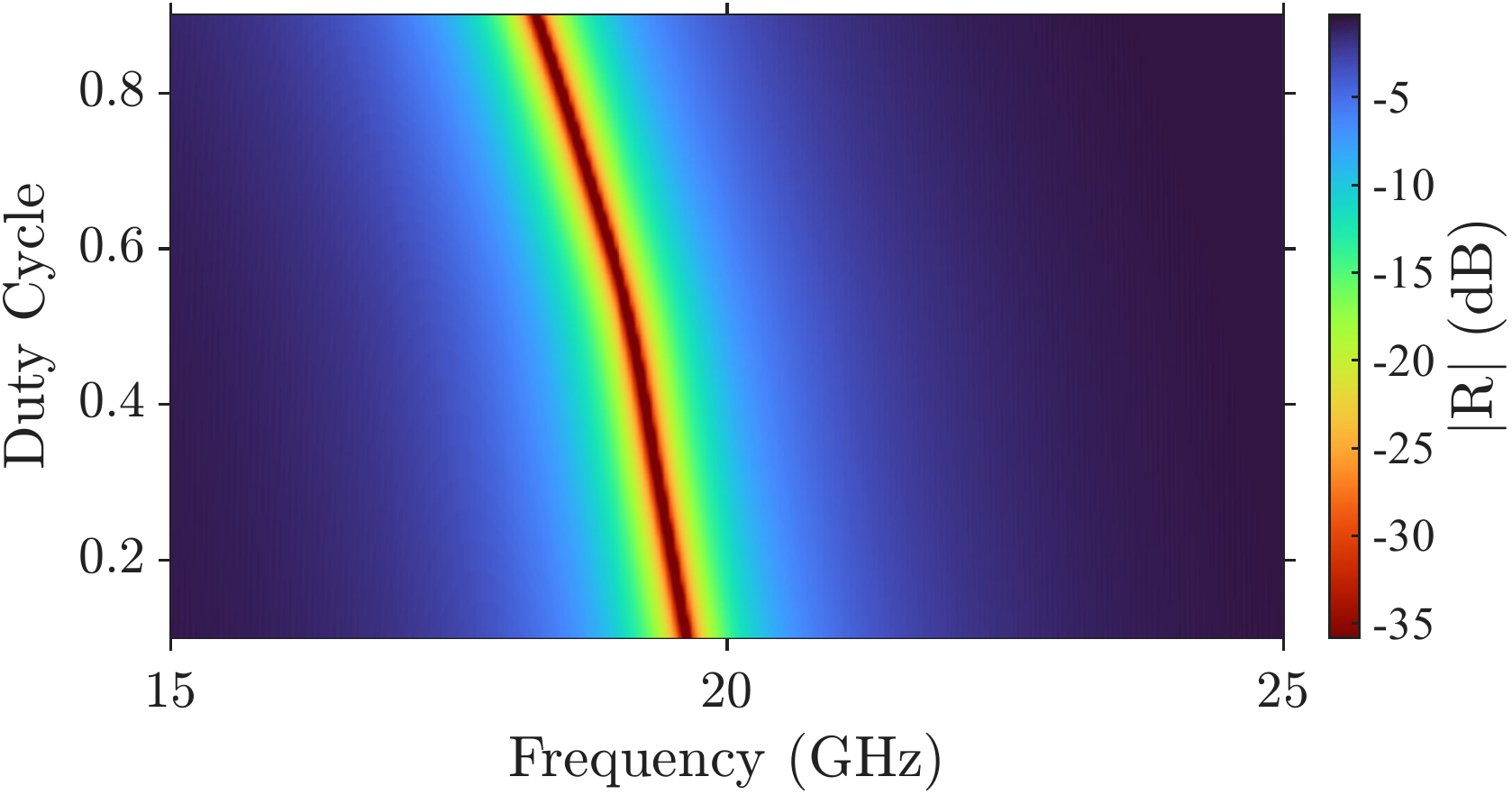}}
    %\subfigure[]{\centering\includegraphics[width=\columnwidth]{figuras/sectionIV/transmision_map.pdf}}
    \caption{Heatmap showing the amplitude of the reflection coefficient, $|R|$, as a function of the frequency and the duty cycle $D$ imposed with the time modulation. A continuous shift of the resonance frequency is observed as $D$ varies. Structure parameters: same as Fig.~\ref{ranura_dividida}.  
    }
    \label{heatmap}
\end{figure}

In a context of resonant cells, time modulation plays the role of an \emph{additional degree of freedom}. To corroborate this, time modulation is assumed to have $f_{\text{M}} = 50\,$MHz ($T_{\text{M}} = 20\,$ns, $\omega_{M} = 2\pi \cdot f_{\text{M}}$). The incident frequency sweeps from $f_{0} = 15\,$GHz to $f_{0} = 25\,$GHz in steps of $50\,$MHz. Notice that for all the frequency points, both $f_{\text{M}}$ and $f_{0}$ are proportional, avoiding scenarios involving the definition of macroperiods \cite{Moreno2023}. In parallel, the duty cycle $D$ is varied from $D = 0$ to $D = 1$, sweeping through all intermediate states. The resulting reflection coefficient is represented in  Fig.~\ref{heatmap}. The vertical axis is left for the duty-cycle variation $D$, whereas the horizontal axis denotes the frequency evolution of $f_0$. 

In Fig.~\ref{heatmap}, the reddish region identifies the frequencies $f_0$ for which the incident wave experiences zero reflection, that is, full transmission, for a given set of temporal parameters. The results show that the duty cycle $D$, which determines the fraction of time the diode remains in the OFF state relative to the ON state, directly controls the resonant response of the space-time metasurface. Specifically, the full-transmission resonance continuously shifts between the resonant frequencies associated with the static OFF and ON states. Unlike in time-invariant scenarios, where only two discrete resonances are observed, the time-modulated metasurface enables continuous tuning of the resonance frequency across the range defined by the static ON and OFF responses, with its exact position governed by the duty cycle.

\begin{figure}[t!]
    \centering    
    \subfigure[]{\includegraphics[width=\columnwidth]{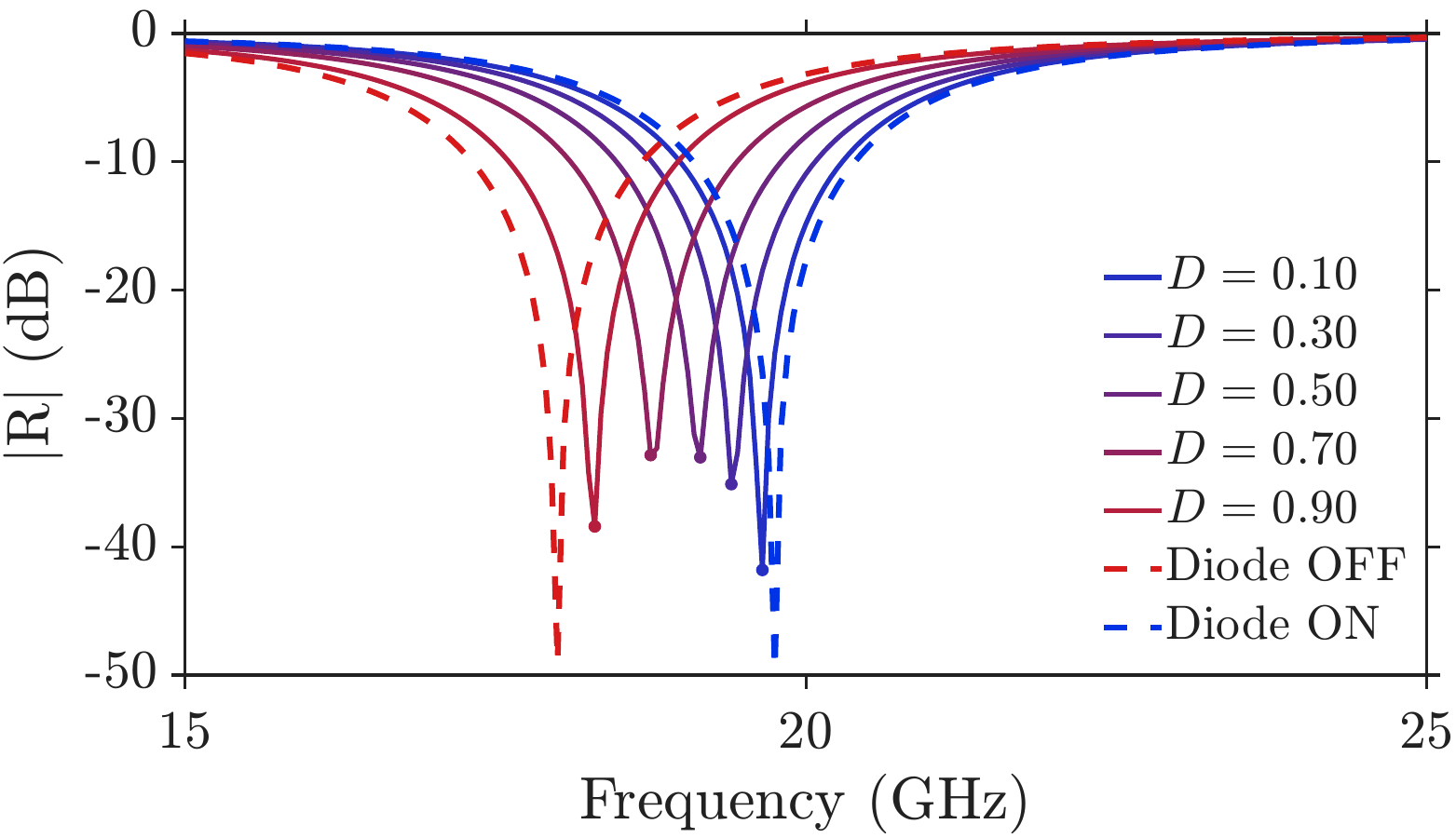}}
    \subfigure[]{\includegraphics[width=\columnwidth]{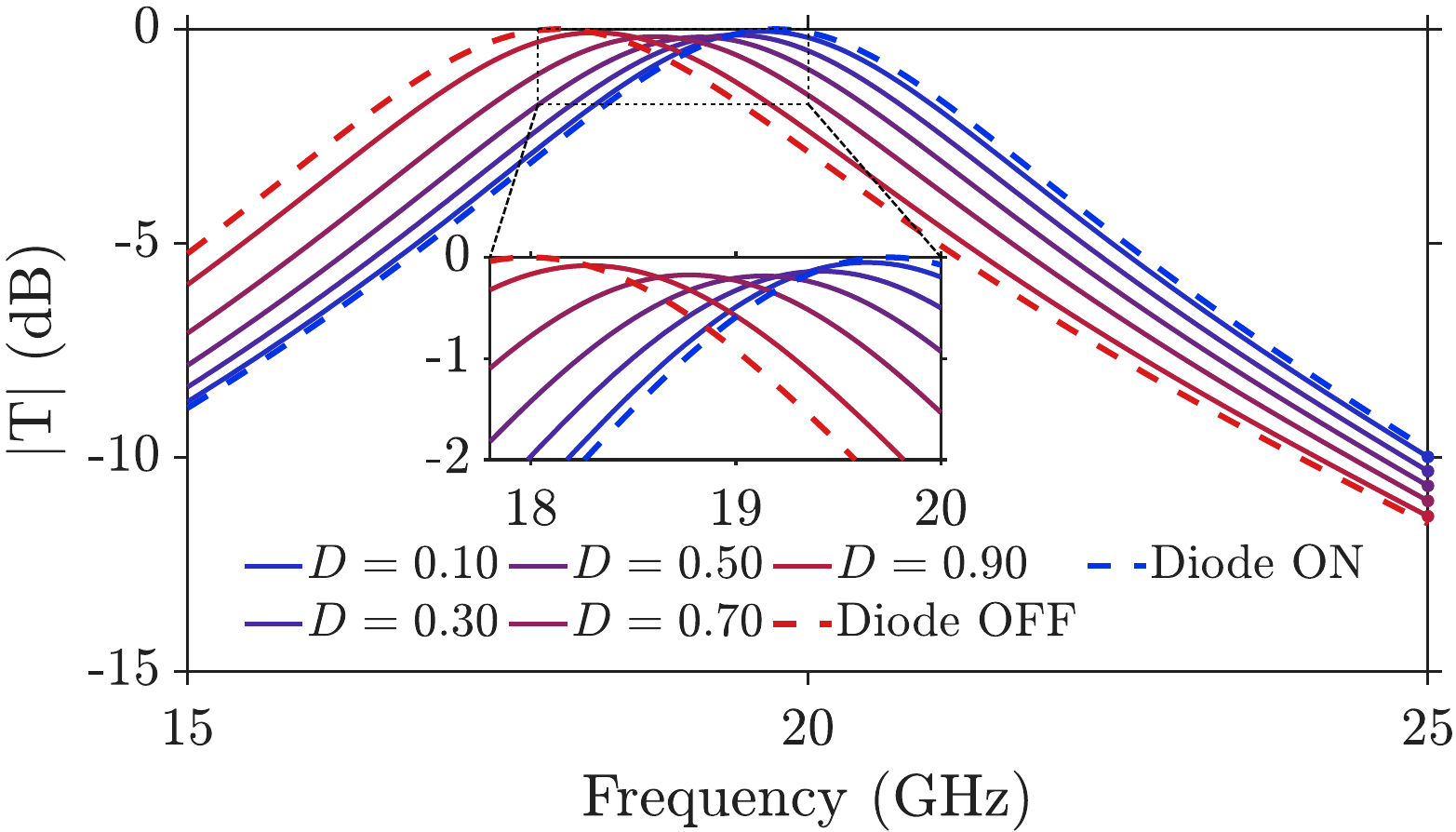}}
    \caption{(a): Reflection coefficient $|R|$ and (b) transmission coefficient $|T|$ versus frequency for different values of the duty cycle, $D$. The figure complements Fig.\ref{heatmap} to illustrate the dynamic tuning of the resonance frequency induced by temporal modulation.}
    \label{curvas_R}
\end{figure}

To further illustrate the spectral evolution, Fig.~\ref{curvas_R} presents the reflection and transmission coefficients for the same scenario shown in Fig.~\ref{heatmap}, together with the stationary ON and OFF responses represented by dashed lines. In the static regime, the resonance shifts from approximately $18$ GHz in the OFF state to nearly $20$ GHz in the ON state. Under spatiotemporal modulation, in which both states alternate periodically, the resonance frequency is directly controlled by the duty cycle $D$. Increasing $D$ shifts the resonance toward lower frequencies, while smaller values move it toward higher frequencies, enabling continuous tuning between the static OFF and ON resonances. This property, unique to space-time modulation, effectively broadens the operational bandwidth from $18$ to $20$ GHz while maintaining full-transmission conditions and offering temporal reconfigurability, achieving a functionality superior to that of conventional multilayer passive structures with a single-layer metasurface.

The underlying physics behind the dynamic tuning relies on the temporal concatenation of the OFF and ON resonant states. The resulting resonant effect manifests an almost full-transmission peak that shifts with $D$. The resonances in steady state (no time modulation) exhibit transmission peaks at 18 GHz and 19.8 GHz when the diode is OFF/ON, respectively. The frequency of the peaks is closely linked to the aperture length, being approximately $w_{\text{y}} \approx \lambda/2$ when the diode is OFF, and $w_{\text{y}1}\approx \lambda/2$ when the diode is ON. This is well-known from the theory of resonators \cite{Munk2000}. The introduction of time modulation imposes periodic sequences of the aperture-length values. In a single periodic cycle, the aperture length is $w_{\text{y}}$ for $0 \le t < DT_{\text{M}}$, and $w_{\text{y}1}$ for $DT_{\text{M}} \le t < T_{\text{M}}$. Since the transmission peaks (reflection nulls) position depends on $D$, the underlying resonant phenomenon is equivalent to that observed with an aperture with effective length $w_{\text{eff}}$ governed by $D$. A very simplified way to estimate this effective aperture length is a weighted average: $w_{\text{eff}} = Dw_{\text{y}} + [1-D] w_{\text{y}1}$. The expression indicates that larger $D$ values result in effective lengths $w_{\text{eff}} \approx w_{\text{y}}$. On the contrary, $w_{\text{eff}} \approx w_{\text{y}1}$ for shorter $D$ values. The intermediate peaks are achieved when $D \in [0.3, 0.7]$ approximately.

Time modulation makes it possible to tune the effective aperture length in a continuous manner, spanning a wide range of values between $w_{\text{y}}$ and $w_{\text{y}1}$. The immediate effect is a high degree of \emph{frequency agility}, as the resonance frequency of the metadevice can be dynamically shifted in real time by simply adjusting the duty cycle of the temporal modulation. This achieves a highly reconfigurable response using only a single-layer structure. In passive platforms, such frequency tuning typically requires complex external circuitry or a set of coupled resonators to cover different operating points. In contrast, space-time modulation introduces a form of \emph{temporal} coupling within the same single resonator, thus avoiding the need for additional physical elements. As a result, devices based on temporal coupling exhibit improved integrability and robustness, while also achieving a lower profile and reduced weight. This technique opens new possibilities for the design of compact, versatile electromagnetic devices with continuously tunable operation.

Following this principle, foreseen  applications focus on truly adaptive RF front-ends, such as software-defined radio (SDR) systems, where duty-cycle control shifts the resonance and eliminates the need for bulky switchable filter banks. By utilizing temporal coupling, a single resonator can be programmed to tune into specific spectral regions, allowing the system to autonomously avoid congested bands or mitigate jamming in real time. This approach significantly reduces system complexity and insertion losses, as the signal path no longer requires multiple switches or redundant physical resonators. Consequently, future 6G networks, where dynamic reconfigurability is fundamental, could greatly benefit from these agile, time-controlled structures. Beyond communications, this framework is also appealing to sensing applications, as the ability to dynamically tune the resonant frequency with high precision enables the detection of specific spectral signatures or shifts in material properties using a single, versatile device.

%---------------------------------------
\section{Conclusions}
This work presents the mathematical derivation of a circuit-based model for analyzing two-dimensional metasurfaces under time modulation. The proposed model accounts for the spatial and temporal periodicity of the unit cell, enabling the representation of the scattered electromagnetic field via a Floquet harmonic expansion. Time modulation is implemented using PIN diodes that alternate between ON and OFF states in a periodic sequence. These diodes are approximated as perfect conductors in the ON state and as open circuits in the OFF state, thereby assuming ultra-fast switching transitions.
Under this framework, two distinct structures are investigated. The first structure emulates a meshed grid subject to time modulation, with particular emphasis on the excitation and analysis of space–time harmonics. The second structure consists of a resonant slot whose resonant characteristics are tuned by adjusting the duty cycle of the time modulation.
Overall, the model provides a quasi-analytical approach for studying time-modulated 2D metasurfaces, highlighting how time modulation serves as an additional degree of freedom to control system behavior and enhance transmission bandwidth.

\begin{acknowledgments}
This work was supported by grant PID2024-155167OA-I00 funded by MCIN/AEI/10.13039/501100011033 and by ERDF/EU. It has also been supported in part by BBVA Foundation's Leonardo Grant for Scientific Research and Cultural Creation 2025, in part by grant Emergia23-00235 funded by Consejería de Universidad, Investigación e Innovación, Junta de Andalucía, in part by grants PID2022-141193OB-I00 and PID2024.157242OB.C44 funded by MCIN/AEI/10.13039/501100011033 and by ERDF/EU, and in part by grant DGP PIDI-2024-00736 funded by Junta de Andalucía. The Foundation takes no responsibility for the opinions, statements, and contents of this project, which are entirely the responsibility of its authors. 
\end{acknowledgments}

\bibliography{./references}% Produces the bibliography via BibTeX.

%--------------------------------
\include{included_supplementary_material}

\end{document}

%% file: included_supplementary_material.tex
%--------------------------------------------------------------------

%SUPLEMENTARY MATERIAL

\setcounter{equation}{0}
\setcounter{figure}{0}
\setcounter{table}{0}
\setcounter{page}{1}
\makeatletter
\renewcommand{\theequation}{S\arabic{equation}}
\renewcommand{\thefigure}{S\arabic{figure}}

\pagebreak

\widetext

\begin{center}
\textrm{\huge Supplementary Material}
\end{center}

\vspace{0.2cm}

\begin{center}
\textrm{\LARGE Time-Controlled Resonances in 2-D Metasurfaces  \\ \vspace{3mm} via Equivalent Circuits}
\end{center}

\begin{center}
\textrm{\Large J. Rafael Sánchez Martínez$^{*}$, Mario Pérez-Escribano$^\dagger$, \\ Antonio Alex-Amor$^\ddagger$, Juan F. Valenzuela-Valdés$^{*}$, \\  Carlos Molero$^{\S}$}
\end{center}

\begin{center}
\textit{$*$ Department of Signal Theory, Telematics and Communications, Centre for Information and Communication Technologies (CITIC-UGR), University of Granada, 18071 Granada, Spain.}
\end{center}

\begin{center}
\textit{$\dagger$Telecommunication Research Institute (TELMA), University of Malaga, E.T.S. Ingeniería de Telecomunicacion, 29010, Málaga}
\end{center}

\begin{center}
\textit{$\ddagger$Department of Electronic and Communication Technology, RFCAS Research Group, Universidad Autónoma de Madrid, 28049 Madrid, Spain}
\end{center}

\begin{center}
\textit{$\S$Department of Electronic and Electromagnetism, Faculty of Physics, University of Seville, 41012, Seville, Spain.}
\end{center}

\section*{Derivation of Equivalent Circuits for Space-time Periodic Structures} \label{Sect-I}

\begin{figure}[t!]
\centering
\subfigure[]{\includegraphics[width=0.55 \columnwidth]{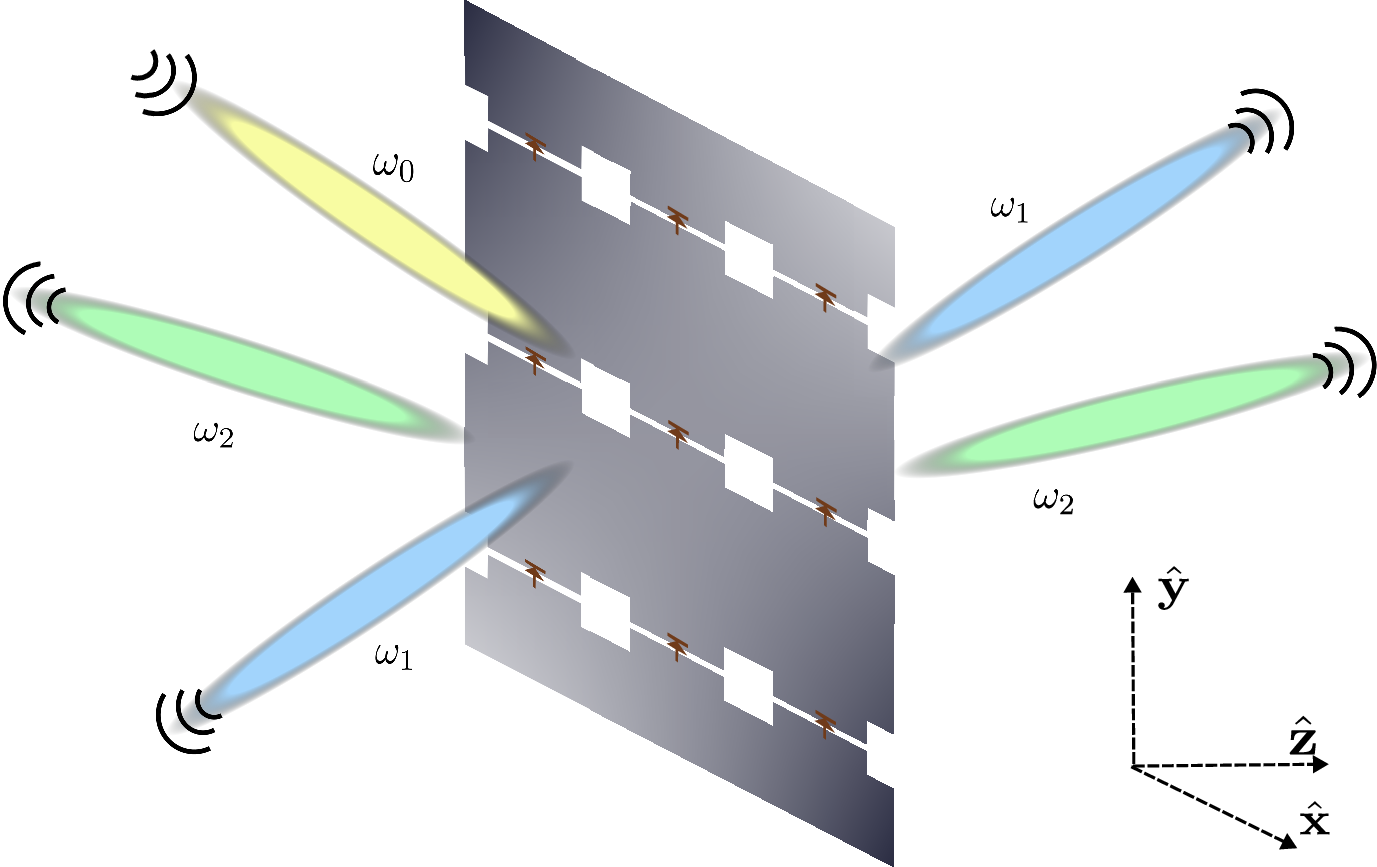}}
\subfigure[]{\includegraphics[width=0.65 \columnwidth]{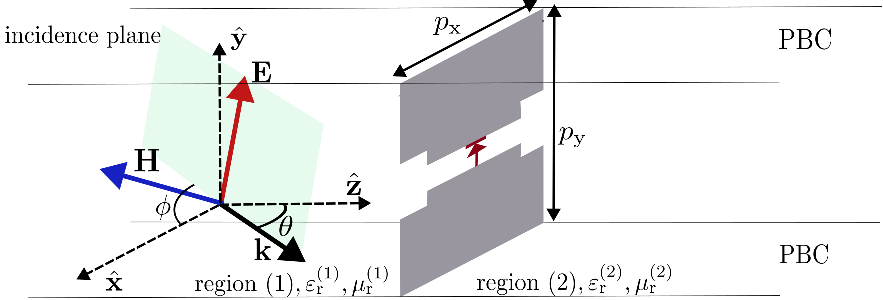}}
\caption{(a) Sketch of a generalized scenario involving a space-time periodic metasurface. The periodic structure is illuminated by a wave with angular frequency $\omega_{0}$. Output beams can have different frequencies in a spatiotemporal scenario, such as $\omega_{1}, \omega_{2}$. (b) Perspective view of the unit cell of the structure. An obliquely incident plane wave illuminates the metasurface to preserve generality in the formulation. The unit cell consists of a thin metallic sheet perforated by slit inclusions, with a diode bridging the metallic regions that define each slit. Two regions, denoted as (1) and (2), are identified and characterized by their respective permittivity and permeability values. The unit cell is enclosed by periodic boundary conditions (PBCs).}
\label{fig:structure}
\end{figure}

The derivation of the equivalent circuit departs from the reduction of the whole periodic structure to a waveguide scenario. Fig.~\ref{fig:structure}(a) shows a general overview of a periodic structure controlled by diodes. The structure is illuminated by a plane wave with angular frequency $\omega_{0}$. Assuming spatio-temporal modulation, the electromagnetic response of the structure manifests as a multi-frequency radiation pattern. The corresponding unit cell is sketched in Fig.~\ref{fig:structure}(b). Oblique incidence is assumed, defined through the incidence angles $\theta$ and $\phi$. The unit-cell analysis is possible thanks to Floquet's theorem. The unit-cell size is delimited by $p_{\text{x}}$ and $p_{\text{y}}$, and special boundaries bound it, called \emph{Periodic Boundary Conditions} (PBCs).

The scenario is equivalent to a waveguide problem with discontinuities inside. From this, three different regions can be distinguished: the incident region (1), where the incident and reflected waves propagate; the transmission region (2), where the transmitted wave travels; and the discontinuity region, where the periodic structure, or the space-time metasurface, is placed.   

The waveguide framework is useful since all the electromagnetic fields propagating within regions (1) and (2) can be described as superpositions of the individual modal solutions of the waveguide. Within the unit-cell realm, the modal solutions are actually Floquet harmonics; thus, the whole electromagnetic field can be described as an \emph{infinite} series expansion of Floquet harmonics. 

%------------------------------------------------------
\subsection*{Time-Invariant Metasurfaces}

The work in \cite{Berral2015} formally defines an individual harmonic from a generalized stationary (time-invariant) scenario; namely, a two-dimensional periodic structure infinitely extended along the $\hat{\mathbf{x}}$ and $\hat{\mathbf{y}}$ directions in space. \textit{In the absence of time modulation}, the mathematical expression of an individual harmonic reads
\begin{equation}\label{E_spatial}
\mathbf{e}_{nm}^{\text{cp/xp}, (v)}(x, y) = \frac{\hat{\mathbf{u}}_{nm}^{\text{cp/xp}}}{\sqrt{p_{\text{x}}p_{\text{y}}}} \text{e}^{\text{j}\omega_{0} t}\,  \text{e}^{-\text{j}( k_{n}x + k_{m}y +  \beta_{nm}^{(v)}z) }\, ,
\end{equation}
where the indices $n$ and $m$ tag the ($n, m$)th harmonic under consideration, $(v)$ refers to the input (1) and output (2) regions, and $\text{cp/xp}$ denotes co-/cross-polarization components. 

Co-polarized harmonics share the vector direction of the incident wave, while cross-polarized harmonics have the orthogonal orientation. Each index, $n$ and $m$, is associated with the spatial periodicities $p_{\text{x}}$ and $p_{\text{y}}$ involved in the problem. The parameters $k_{n}$ and $k_{m}$, 
\begin{align}
k_{n} &= k_{\text{x}} + \frac{2 n \pi }{p_{\text{x}}} \hspace{5 mm} n \in \mathbb{Z} \\ 
k_{m} &= k_{\text{y}} + \frac{2 m \pi }{p_{\text{y}}} \hspace{5 mm} m \in \mathbb{Z}
\end{align}
are the transverse cutoff wavenumbers of the ($n,m$)th Floquet harmonic. Notice that these entities depend on $k_{\text{x}}$ and $k_{\text{y}}$, both the transverse wavenumbers of the incident wave,
\begin{align}
k_{\text{x}} = \sqrt{\varepsilon_{\text{r}}^{(1)} \mu_{\text{r}}^{(1)}} \frac{\omega_{0}}{c} \cos(\phi) \sin(\theta) \\
k_{\text{y}} = \sqrt{\varepsilon_{\text{r}}^{(1)} \mu_{\text{r}}^{(1)}} \frac{\omega_{0}}{c} \sin(\phi) \sin(\theta) ,
\end{align}
with $\theta$ being the elevation and $\phi$ the  azimuth in spherical coordinates (incidence angles). The coefficients $\varepsilon_{\text{r}}^{(1)}$ and  $\mu_{\text{r}}^{(1)}$ are the relative permittivity and permeability in medium (1),  $\omega_{0}$ is the angular frequency of the incident wave, and $c$ the speed of light. 

\begin{figure}[]
\centering
\subfigure[]{\includegraphics[width=0.3 \columnwidth]{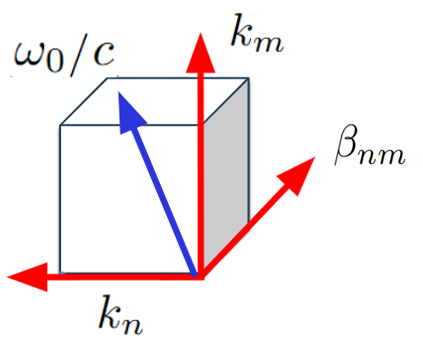}} \hspace{6 mm}
\subfigure[]{\includegraphics[width=0.3 \columnwidth]{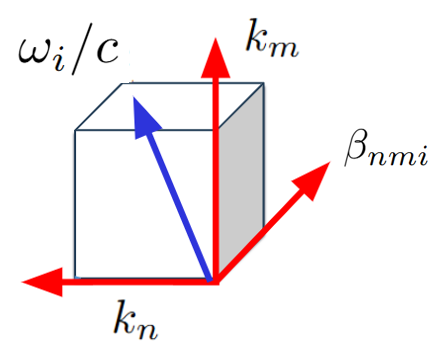}}
\caption{(a) Graphical representation of a ($n,m$)th-order spatial harmonic. (b) Graphical representation of an $n,m,i)$th-order spatiotemporal harmonic. }
\label{fig:comparison}
\end{figure}

The longitudinal component of the wavevector, $\beta_{nm}^{(v)}$, is expressed in the stationary time-invariant case as 
\begin{equation}
\beta_{nm}^{(v)} = \sqrt{\varepsilon_{\text{r}}^{(v)} \mu_{\text{r}}^{(v)} \frac{\omega_{0}^2}{c^2} - k_{n}^{2} - k_{m}^2}.
\end{equation}
The harmonic vector information is captured by $\hat{\mathbf{u}}_{nm}^{\text{cp/xp}}$, defined in \cite{Berral2015} as
\begin{align}
\hat{\mathbf{u}}_{nm}^{\text{TM}} &= \frac{k_{n}\hat{\mathbf{x}} + k_{m}\hat{\mathbf{y}}}{k_{n}^2 + k_{m}^2} \\
\hat{\mathbf{u}}_{nm}^{\text{TE}} &= \frac{-k_{n}\hat{\mathbf{y}} + k_{m}\hat{\mathbf{x}}}{k_{n}^2 + k_{m}^2}\,. 
\end{align}
The superscript TM/TE refers to the polarization of the harmonic. If the incident one is of TE-nature, this polarization becomes the co-polarized one (cp), and TM becomes the cross-polarized one (xp). These labels are interchanged when the incident polarization is TM. A schematic representation of the vector distribution of a harmonic is shown in Fig.~\ref{fig:comparison}(a). The whole wavevector, whose modulus is $\omega_{0}/c$, is the result of the vector sum between $k_{n}, k_{m}$  and $\beta_{nm}^{(v)} $. 
 \\

Once the individual ($n,m$)th harmonic is defined, we can analytically describe the whole electromagnetic field in regions (1) and (2) as a Floquet-Bloch series:
\begin{align}\label{E1}
\mathbf{E}^{(1)}(x, y, z, t) &= \left\{1\cdot \mathbf{e}_{00}^{\text{cp}, (1)} + \displaystyle\sum_{n = -\infty}^{n = \infty} \displaystyle\sum_{m = - \infty}^{m = \infty} \big[E_{nm}^{\text{cp}, (1)} \mathbf{e}_{nm}^{\text{cp}, (1)}  + E_{nm}^{\text{xp}, (1)} \mathbf{e}_{nm}^{\text{xp}, (1)} \big]\right\} \text{e}^{\text{j}\omega_{0}t}\\
\label{E2}
\mathbf{E}^{(2)}(x, y, z, t) &=  \left\{\displaystyle\sum_{n = -\infty}^{n = \infty} \displaystyle\sum_{m = - \infty}^{m = \infty} \big[E_{nm}^{\text{cp}, (2)} \mathbf{e}_{nm}^{\text{cp}, (2)}  + E_{nm}^{\text{xp}, (2)} \mathbf{e}_{nm}^{\text{xp}, (2)} \big]\right\} \text{e}^{\text{j} \omega_{0}t},
\end{align}
%\textcolor{red}{CARLOS: Si lo ponemos así, hay que quitar la exponencial de la expresión en (1). Antonio: quizás es entonces más fácil ponerla solo en (1) y no repetirla tantas veces en el resto de ecuaciones}. 
where we have used the same description as \cite{Berral2015} but with a slightly modified notation. The term out of the summation in \eqref{E1} denotes the incident wave, with amplitude unity. From the infinite series, the co-polarized amplitude fields with orders $n = m = 0$, $E_{00}^{\text{cp}, (1)}, E_{00}^{\text{cp}, (2)}$, are recognized as the reflection $R$ and transmission $T$ coefficients, also identified as $E_{00}^{\text{cp}, (1)} = R$, $E_{00}^{\text{cp}, (2)} = T$. Similarly, the cross-polarized counterparts are included in $E_{00}^{\text{xp}, (1)}$ and $E_{00}^{\text{xp}, (2)}$. These amplitudes are unknown, as well as the rest of the amplitude coefficients associated with the highest-order harmonics $E_{nm}^{\text{cp/xp}, (v)}$. 

A similar field expansion is carried out for the magnetic field in regions (1) and (2), 
\begin{multline}\label{H1}
\mathbf{H}^{(1)}(x, y, z, t) = \Bigg\{ Y_{00}^{\text{cp}, (1)}\mathbf{e}_{00}^{\text{cp}, (1)} \Bigg. \\ \Bigg. - \displaystyle\sum_{n = -\infty}^{n = \infty} \displaystyle\sum_{m = - \infty}^{m = \infty} \big[Y_{nm}^{\text{cp}, (1)}E_{nm}^{\text{cp}, (1)} \mathbf{e}_{nm}^{\text{cp}, (1)}  + Y_{00}^{\text{xp}, (1)}E_{nm}^{\text{xp}, (1)} \mathbf{e}_{nm}^{\text{xp}, (1)} \big] \Bigg\}  \cdot \text{e}^{\text{j} \omega_{0} t} 
\end{multline}
\begin{equation}\label{H2}
\mathbf{H}^{(2)}(x, y, z, t) = \left\{\displaystyle\sum_{n = -\infty}^{n = \infty} \displaystyle\sum_{m = - \infty}^{m = \infty} \big[Y_{nm}^{\text{cp}, (2)}E_{nm}^{\text{cp}, (2)} \mathbf{e}_{nm}^{\text{cp}, (2)}  + Y_{00}^{\text{xp}, (2)}E_{nm}^{\text{xp}, (2)} \mathbf{e}_{nm}^{\text{xp}, (2)} \big] \right\} \cdot \text{e}^{\text{j} \omega_{0} t} ,
\end{equation}
where now the admittance coefficients appear:
\begin{align}
Y_{nm}^{\text{TE}, (v)} &= \frac{\beta_{nm}^{(v)}}{ \mu_{0} \mu_{\text{r}}^{(v)} \omega_{0}} \\
Y_{nm}^{\text{TM}, (v)} &= \frac{\varepsilon_{0}\varepsilon_{\text{r}}^{(v)} \omega_{0}}{\beta_{nm}^{(v)}} \hspace{5mm} v = 1,2 \,.
\end{align}
The time evolution is harmonic in this case, as stated by $\text{e}^{\text{j} \omega_{0} t}$, due to the absence of time modulation. In these circumstances, the time evolution can be removed from the calculations, and the problem can be solved solely in terms of spatial variables. 

%------------------------------------------------------
\subsection*{Time-Varying Metasurfaces}

The introduction of time modulation leads to a \emph{non-harmonic temporal evolution} of the field distributions in regions (1) and (2). The whole electromagnetic system can be described as the interaction between two different entities vibrating with distinct frequencies: the incident field, vibrating with $\omega_{0} = 2\pi/T_{0}$; and the periodic structure, modulated by $\omega_{\text{M}} = 2\pi/T_{\text{M}}$. The resulting radiation fields are somehow identified as an infinite summation of individual fields vibrating with the inter-modulation products $\omega_{0} \pm \omega_{\text{M}}$,  $\omega_{0} \pm 2 \omega_{\text{M}}$, etc. Each term can be interpreted as the contribution of an individual \emph{Floquet temporal harmonic}, in close analogy with the case of purely spatial modulations. From this perspective, a \emph{complete} individual harmonic incorporates both spatial and temporal periodicity, and therefore the expression in \eqref{E_spatial} can be generalized as follows:
\begin{equation}\label{E_spatial}
\mathbf{e}_{nmi}^{\text{cp/xp}, (v)}(x, y, t) = \frac{\hat{\mathbf{u}}_{nm}^{\text{cp/xp}}}{\sqrt{p_{\text{x}}p_{\text{}y}T_{\text{M}}}} \text{e}^{\text{j}\omega_{i} t}\,  \text{e}^{-\text{j}( k_{n}x + k_{m}y +  \beta_{nmi}^{(v)}z) }\, ,
\end{equation}
where the index $i$ (it should not be confused with the imaginary unit, which is here represented as $\mathrm{j}$) is due to the time periodicity, and
\begin{equation}
\omega_{i} = \omega_{0} + i\omega_{\text{M}} \hspace{5mm} i \in \mathbb{Z} \,.
\end{equation}
The time period $T_{\text{M}}$ corresponds to the so-called \emph{macroperiod}, defined as the interval over which the two processes evolving at $\omega_{0}$ and $\omega_{\text{M}}$, respectively, simultaneously complete an integer number of cycles \cite{Moreno2023}. 

Accordingly, the propagation constant can now be expressed as  
\begin{equation}
\beta_{nmi}^{(v)} = \sqrt{\varepsilon_{\text{r}}^{(v)} \mu_{\text{r}}^{(v)} \frac{\omega_{i}^2}{c^2} - k_{n}^2 - k_{m}^2}, \hspace{5mm} v = 1, 2\,,
\end{equation}
thus the vector representation of an individual $(n,m,i)$th spatiotemporal harmonic is that in Fig.~\ref{fig:comparison}(b).

The electric fields in regions (1) and (2) are now expressed as follows:
\begin{align}\label{E1t}
\mathbf{E}^{(1)}(x, y, z, t) = 1\cdot \mathbf{e}_{000}^{\text{cp}, (1)} + \displaystyle\sum_{n = -\infty}^{\infty} \displaystyle\sum_{m = - \infty}^{\infty} \displaystyle\sum_{i = -\infty}^{\infty} \big[E_{nmi}^{\text{cp}, (1)} \mathbf{e}_{nmi}^{\text{cp}, (1)}  + E_{nm}^{\text{xp}, (1)} \mathbf{e}_{nmi}^{\text{xp}, (1)} \big]\\
\label{E2t}
\mathbf{E}^{(2)}(x, y, z, t) =  \displaystyle\sum_{n = -\infty}^{\infty} \displaystyle\sum_{m = - \infty}^{\infty} \displaystyle\sum_{i = - \infty} ^{\infty} \big[E_{nmi}^{\text{cp}, (2)} \mathbf{e}_{nmi}^{\text{cp}, (2)}  + E_{nmi}^{\text{xp}, (2)} \mathbf{e}_{nmi}^{\text{xp}, (2)} \big] ,
\end{align}
whereas the magnetic field reads
\begin{multline}\label{H1t}
\mathbf{H}^{(1)}(x, y, z, t) = Y_{000}^{\text{cp}, (1)}\mathbf{e}_{000}^{\text{cp}, (1)}  - \displaystyle\sum_{n = -\infty}^{\infty} \displaystyle\sum_{m = - \infty}^{\infty} \displaystyle\sum_{i = -\infty}^{\infty} \big[Y_{nmi}^{\text{cp}, (1)}E_{nmi}^{\text{cp}, (1)} \mathbf{e}_{nmi}^{\text{cp}, (1)}  + Y_{nmi}^{\text{xp}, (1)}E_{nmi}^{\text{xp}, (1)} \mathbf{e}_{nm}^{\text{xp}, (1)} \big]   
\end{multline}
\begin{equation}\label{H2t}
\mathbf{H}^{(2)}(x, y, z, t) = \displaystyle\sum_{n = -\infty}^{\infty} \displaystyle\sum_{m = - \infty}^{\infty} 
\displaystyle\sum_{i = -\infty}^{\infty} \big[Y_{nmi}^{\text{cp}, (2)}E_{nmi}^{\text{cp}, (2)} \mathbf{e}_{nmi}^{\text{cp}, (2)}  + Y_{nmi}^{\text{xp}, (2)}E_{nmi}^{\text{xp}, (2)} \mathbf{e}_{nmi}^{\text{xp}, (2)} \big] 
\end{equation}

\begin{align}
Y_{nmi}^{\text{TE}, (v)} &= \frac{\beta_{nmi}^{(v)}}{ \mu_{0} \mu_{\text{r}}^{(v)} \omega_{\text{i}}} \\
Y_{nmi}^{\text{TM}, (v)} &= \frac{\varepsilon_{0}\varepsilon_{\text{r}}^{(v)} \omega_{\text{i}}}{\beta_{nmi}^{(v)}} \hspace{5mm} v = 1, 2 \,.
\end{align}

\subsection*{Computation of the Unknowns. Boundary Conditions}

To obtain the values of the unknowns, the Floquet coefficients $E_{nmi}^{\text{cp/xp, (v)}}$, we need additional information about the fields in the discontinuity region. For simplicity, the discontinuity region is placed at $z = 0$. The electric field there is denoted as $\mathbf{E}_{\text{s}}(x, y, t)$ and directly depends on the geometry of the scatterer (spatial dependence along $\hat{\mathbf{x}}$ and $\hat{\mathbf{y}}$), the polarization of the incident wave, and the modulation period $T_{\text{M}}$. Time modulation can be implemented using a reconfigurable element biased by a periodic source. The example in Fig.~\ref{fig:structure} is modulated by a diode, for instance.

Importantly, for the derivation of the circuit model, \textit{the knowledge, a priori, of the properties of $\mathbf{E}_{\text{s}}(x, y, t)$ is crucial}. Some previous works have focused on this key point and solved the problem for many geometrical forms of the scatterers  \cite{Alex2021, Molero2021}. Two main boundary conditions are subsequently established, where $\mathbf{E}_{\text{s}}(x, y, t)$ plays a fundamental role.

On the one hand, and assuming \emph{instantaneous} time switching in the reconfigurable element controlling time modulation, the continuity of the electric field across the discontinuity must be satisfied:
\begin{align}\label{Econt1}
\mathbf{E}^{(1)}(x, y, z = 0, t) &= \mathbf{E}_{\text{s}}(x, y, t) \\ 
\label{Econt2} \mathbf{E}^{(2)}(x, y, z = 0, t) &= \mathbf{E}_{\text{s}}(x, y, t)\,.
\end{align}
Transient phenomena are not considered in this study, as they are expected to be irrelevant at the frequencies of interest, primarily RF, microwave, and millimeter-wave frequencies \cite{ZapataTV2024}.

By focusing on the equality in \eqref{Econt1}, and introducing \eqref{E1t}, we obtain 
\begin{equation}
\mathbf{e}_{000}^{\text{cp}, (1)} + \displaystyle\sum_{n = -\infty}^{\infty} \displaystyle\sum_{m = - \infty}^{\infty} \displaystyle\sum_{i = -\infty}^{\infty} \big[E_{nmi}^{\text{cp}, (1)} \mathbf{e}_{nmi}^{\text{cp}, (1)}  + E_{nm}^{\text{xp}, (1)} \mathbf{e}_{nmi}^{\text{xp}, (1)} \big] = \mathbf{E}_{\text{s}}(x, y, t)\, . 
\end{equation}
Due to the orthogonality of the different harmonics, it is straightforward to show that the following relationship holds for the incident harmonic
\begin{equation}
[1 + R] = \displaystyle\int_{p_{\text{x}}} \displaystyle\int_{p_{\text{y}}} \displaystyle\int_{T_{\text{M}}} \mathbf{E}_{\text{s}}(x, y, t) \cdot [\mathbf{e}_{000}^{\text{cp},(1)}]^{*}  \text{d}x \text{d}y  \text{d}t, 
\end{equation}
which, after some manipulations, reads 
\begin{equation}\label{E000co}
[1 + R] = \frac{1}{\sqrt{p_{\text{x}} p_{\text{y}} T_{\text{M}} }}\displaystyle\int_{p_{\text{x}}} \displaystyle\int_{p_{\text{y}}} \displaystyle\int_{T_{\text{M}}} \mathbf{E}_{\text{s}}(x, y, t) \cdot \hat{\mathbf{u}}_{000}^{\text{cp}} \text{e}^{-\text{j}\omega_{0} t}\,  \text{e}^{\text{j}( k_{0}x + k_{0}y)} \text{d}x \text{d}y \text{d}t \,.
\end{equation}
By similar arguments, 
\begin{align}\label{E000cross}
E_{000}^{\text{xp}, (1)} &= \frac{1}{\sqrt{p_{\text{x}} p_{\text{y}} T_{\text{M}} }}\displaystyle\int_{p_{\text{x}}} \displaystyle\int_{p_{\text{y}}} \displaystyle\int_{T_{\text{M}}} \mathbf{E}_{\text{s}}(x, y, t) \cdot \hat{\mathbf{u}}_{000}^{\text{xp}} \text{e}^{-\text{j}\omega_{0} t}\,  \text{e}^{\text{j}( k_{0}x + k_{0}y)} \text{d}x \text{d}y \text{d}t, \\
\label{Enmico}
E_{nmi}^{\text{cp}, (1)} &= \frac{1}{\sqrt{p_{\text{x}} p_{\text{y}} T_{\text{M}} }}\displaystyle\int_{p_{\text{x}}} \displaystyle\int_{p_{\text{y}}} \displaystyle\int_{T_{\text{M}}} \mathbf{E}_{\text{s}}(x, y, t) \cdot \hat{\mathbf{u}}_{nmi}^{\text{cp}} \text{e}^{-\text{j}\omega_{i} t}\,  \text{e}^{\text{j}( k_{n}x + k_{m}y)} \text{d}x \text{d}y \text{d}t, \\
\label{Enmicross}
E_{nmi}^{\text{xp}, (1)} &= \frac{1}{\sqrt{p_{\text{x}} p_{\text{y}} T_{\text{M}} }}\displaystyle\int_{p_{\text{x}}} \displaystyle\int_{p_{\text{y}}} \displaystyle\int_{T_{\text{M}}} \mathbf{E}_{\text{s}}(x, y, t) \cdot \hat{\mathbf{u}}_{nmi}^{\text{xp}} \text{e}^{-\text{j}\omega_{i} t}\,  \text{e}^{\text{j}( k_{n}x + k_{m}y)} \text{d}x \text{d}y \text{d}t,
\end{align}
where it can be inferred that the amplitude of each of the harmonics is quantified in terms of the $(n,m,i)$th-order Fourier transform of the field $\mathbf{E}_{\text{s}}(x, y, t)$. The limits of integration ${p_{\text{x}}}$, ${p_{\text{y}}}$, and ${T_{\text{M}}}$ indicate that the integration is confined within each period; namely, $x\in[-{p_{\text{x}}}/2, {p_{\text{x}}}/2]$, $y\in[-{p_{\text{y}}}/2, {p_{\text{y}}}/2]$, and $t\in[-{T_{\text{M}}}/2, {T_{\text{M}}}/2]$. %\red{Antonio: he añadido esta última frase, cómo la véis?}

In addition, all the higher-order harmonic amplitudes can be expressed as a function of the reflection coefficient $R$. By dividing \eqref{E000cross}, \eqref{Enmico}, \eqref{Enmicross} by \eqref{E000co}, one obtains:
\begin{align}
E_{000}^{\text{xp}, (1)} &= [1 + R] \underbrace{\frac{{\displaystyle\int_{p_{\text{x}}} \displaystyle\int_{p_{\text{y}}} \displaystyle\int_{T_{\text{M}}} \mathbf{E}_{\text{s}}(x, y, t) \cdot \hat{\mathbf{u}}_{000}^{\text{xp}} \text{e}^{-\text{j}\omega_{0} t}\,  \text{e}^{\text{j}( k_{0}x + k_{0}y)} \text{d}x \text{d}y \text{d}t}}{\displaystyle\int_{p_{\text{x}}} \displaystyle\int_{p_{\text{y}}} \displaystyle\int_{T_{\text{M}}} \mathbf{E}_{\text{s}}(x, y, t) \cdot \hat{\mathbf{u}}_{000}^{\text{cp}} \text{e}^{-\text{j}\omega_{0} t}\,  \text{e}^{\text{j}( k_{0}x + k_{0}y)} \text{d}x \text{d}y \text{d}t}}_{N_{000}^{\text{xp}}} \\
E_{nmi}^{\text{cp}, (1)} &= [1 + R] \underbrace{\frac{{\displaystyle\int_{p_{\text{x}}} \displaystyle\int_{p_{\text{y}}} \displaystyle\int_{T_{\text{M}}} \mathbf{E}_{\text{s}}(x, y, t) \cdot \hat{\mathbf{u}}_{nmi}^{\text{cp}} \text{e}^{-\text{j}\omega_{i} t}\,  \text{e}^{\text{j}( k_{n}x + k_{m}y)} \text{d}x \text{d}y \text{d}t}}{\displaystyle\int_{p_{\text{x}}} \displaystyle\int_{p_{\text{y}}} \displaystyle\int_{T_{\text{M}}} \mathbf{E}_{\text{s}}(x, y, t) \cdot \hat{\mathbf{u}}_{000}^{\text{cp}} \text{e}^{-\text{j}\omega_{0} t}\,  \text{e}^{\text{j}( k_{0}x + k_{0}y)} \text{d}x \text{d}y \text{d}t}}_{N_{nmi}^{\text{cp}}} \\ 
E_{nmi}^{\text{xp}, (1)} &= [1 + R] \underbrace{\frac{{\displaystyle\int_{p_{\text{x}}} \displaystyle\int_{p_{\text{y}}} \displaystyle\int_{T_{\text{M}}} \mathbf{E}_{\text{s}}(x, y, t) \cdot \hat{\mathbf{u}}_{nmi}^{\text{xp}} \text{e}^{-\text{j}\omega_{i} t}\,  \text{e}^{\text{j}( k_{n}x + k_{m}y)} \text{d}x \text{d}y \text{d}t}}{\displaystyle\int_{p_{\text{x}}} \displaystyle\int_{p_{\text{y}}} \displaystyle\int_{T_{\text{M}}} \mathbf{E}_{\text{s}}(x, y, t) \cdot \hat{\mathbf{u}}_{000}^{\text{cp}} \text{e}^{-\text{j}\omega_{0} t}\,  \text{e}^{\text{j}( k_{0}x + k_{0}y)} \text{d}x \text{d}y \text{d}t}}_{N_{nmi}^{\text{xp}}} ,
\end{align}
where the terms $N_{nmi}^{\text{cp/xp}}$ have been defined, and will later be interpreted as complex-valued transformers with turn ratio $N_{nmi}:1$ from the circuit standpoint. 
By applying the same procedure for medium (2) and the expression in \eqref{Econt2}, it is straightforward to deduce that
\begin{align}
[1 + R] &= T \\
E_{000}^{\text{xp}, (1)} &= E_{000}^{\text{xp}, (2)} \\
E_{nmi}^{\text{cp}, (1)} &= E_{nmi}^{\text{cp}, (2)} \\
E_{nmi}^{\text{xp}, (1)} &= E_{nmi}^{\text{xp}, (2)}.
\end{align}
Thus, the superscripts 
$(1)$ and $(2)$ will be omitted hereafter.

\begin{figure}[t!]
  \centering
    \subfigure[]{\includegraphics[width= 0.6\columnwidth]{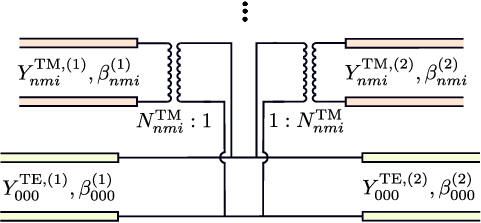}}
    \subfigure[]{\includegraphics[width= 0.6\columnwidth]{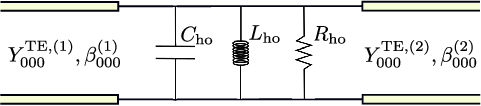}}
  \caption{(a): Equivalent circuit. TE incidence is assumed. The picture shows only a single higher-order harmonic of TM nature. The dots symbolize an infinite connection of lines, encompassing both TE and TM nature. (b): Reduced version of the circuit, where the parallel connection of a capacitor, an inductor, and a resistor can emulate the presence of the higher-order harmonics.}
  \label{fig:circuit}
\end{figure}

A second boundary condition, involving the continuity of Poynting vector across the discontinuity, is imposed:
\begin{equation}\label{cond2}
\mathbf{E}_{\text{s}}(x, y, z) \times \mathbf{H}^{(1)}(x, y, z = 0, t) = \mathbf{E}_{\text{s}}(x, y, z) \times \mathbf{H}^{(2)}(x, y, z = 0, t)\,.
\end{equation}
In a similar way as in the previous condition, we introduce \eqref{H1} and \eqref{H2} in \eqref{cond2}, and make use of the orthogonality among harmonics. 
This condition, as well as that in \eqref{Econt1}-\eqref{Econt2}, is imposed in temporal regions where the discontinuity remains invariant. We are assuming instantaneous switching. At the switching instants, the fields to be continuous are the electric and magnetic flux $\mathbf{D}(t)$ and $\mathbf{B}(t)$. In the temporal regions outside the material transition, the continuity of power can be established. Notice that this constitutes an approximation of our model, which is supposed to work well when the periodicity of time modulation $T_{\text{M}}$ is much longer than the periodicity of the incident wave $T_{0}$. Thus, transient times due to the transition can be neglected. This strategy has been adopted in some works with experimental verification, such as those related to space-time coding metasurfaces \cite{Cui2014, ZhanCui2018}. 

After some calculations, we obtain, 
\begin{multline}
(1 - R)Y_{000}^{\text{cp}, (1)} -(1 + R)Y_{000}^{\text{xp}, (2)} - (1 + R)\displaystyle\sum_{\forall n, m, i}^{'} \big[ Y_{nmi}^{\text{cp}, (1)}|N_{nmi}^{\text{cp}}|^2 + Y_{nmi}^{\text{xp}, (1)}|N_{nmi}^{\text{xp}}|^2\big]  \\
= (1 + R)Y_{000}^{\text{cp}, (2)} + (1 + R)Y_{000}^{\text{xp}, (2)} + (1 + R)\displaystyle\sum_{\forall n, m, i}^{'} \big[ Y_{nmi}^{\text{cp}, (2)}|N_{nmi}^{\text{cp}}|^{2} + Y_{nmi}^{\text{xp}, (2)}|N_{nmi}^{\text{xp}}|^{2}\big] .
\end{multline}
Thus, rearranging, a formal expression for the reflection coefficient is readily found, 
%\red{Antonio: es posible que en la segunda línea de la ec. (40) los superíndices sean (2) en vez de (1)? O quizás son (1) y (2) combinados? Revisar} 
\begin{equation}\label{reflection}
R = \frac{Y_{000}^{\text{cp}, (1)} - Y_{000}^{\text{cp}, (2)} - Y_{\text{eq}}}{Y_{000}^{\text{cp}, (1)} + Y_{000}^{\text{cp}, (2)} + Y_{\text{eq}}},
\end{equation}
with 
\begin{equation}
Y_{\text{eq}} = \displaystyle\sum_{\forall n,m,i}^{'} [Y_{nmi}^{\text{cp}, (1)} + Y_{nmi}^{\text{cp}, (2)}] |N_{nmi}^{\text{cp}}|^2 + \displaystyle\sum_{\forall n,m,i}  [Y_{nmi}^{\text{xp}, (1)} + Y_{nmi}^{\text{xp}, (2)}] |N_{nmi}^{\text{xp}}|^2,
\end{equation}
where the prime symbol ($'$) above the sums indicates that the summation takes any combination of $n, m, i$ except for the incident harmonic $n = m = i = 0$. 

The infinite summation in \eqref{reflection} is identified as a parallel-connection of admittances from a circuit standpoint. Each of the admittances admit to be described as a $(n, m, i)$th-order transmission line with characteristic admittance $Y_{nmi}^{\text{cp/xp},(v)}$ and propagation constant $\beta_{nmi}^{(v)}$. Fig.~\ref{fig:circuit}(a) shows a schematic representation of the resulting topology. Without loss of generality, TE incidence is assumed. This leads to the excitation of co-polarized TE harmonics and cross-polarized TM harmonics. Each of these harmonics is represented by an individual transmission line with admittance $Y_{nmi}^{\text{TE/TM},(v)}$ and propagation constant $\beta_{nmi}^{(v)}$. In addition, it is connected to the transmission lines of the fundamental harmonic through a transformer with turn ratio $N_{nmi}^{\text{cp/xp}}:1$. A simplified version of the circuit as such in Fig.~\ref{fig:circuit}(b) is admitted by assuming that: all TM evanescent harmonics contribute capacitively; all TE evanescent harmonics contribute inductively; and all the TM/TE propagating harmonics carry power, thus they all can be involved in a global resistor.

\section*{\label{sec:Sect3} Previous cases supporting the model}
As mentioned in the introduction of the main text, as well as in the discussion in Sect. II-B, there is a significant lack of full-wave solvers to account for space-time problems. Specifically, for 2-D platforms under time modulation. This limitation motivates the use of partial validation strategies to assess the model's feasibility. While a partial validation does not verify the entire model, a set of complementary partial validations can provide sufficient robustness and confidence in the proposed formulation.

A first partial validation comes from the assumption that the transition from metal to air and vice versa can be considered almost instantaneous. This situation is perfectly given when the time modulation is substantially smaller than the frequency of the incident wave, say tens of MHz versus few GHz. Under this small-modulation assumption, transient regimes are negligible. The circuit topologies for the ON/OFF states of the diodes (Fig.~3 of the main manuscript) tend to become short/open circuits, respectively. This assumption has also been experimentally verified in \cite{Cui2014, ZhanCui2018}. 

\begin{figure}[t!]
    \centering    \includegraphics[width=0.89\columnwidth]{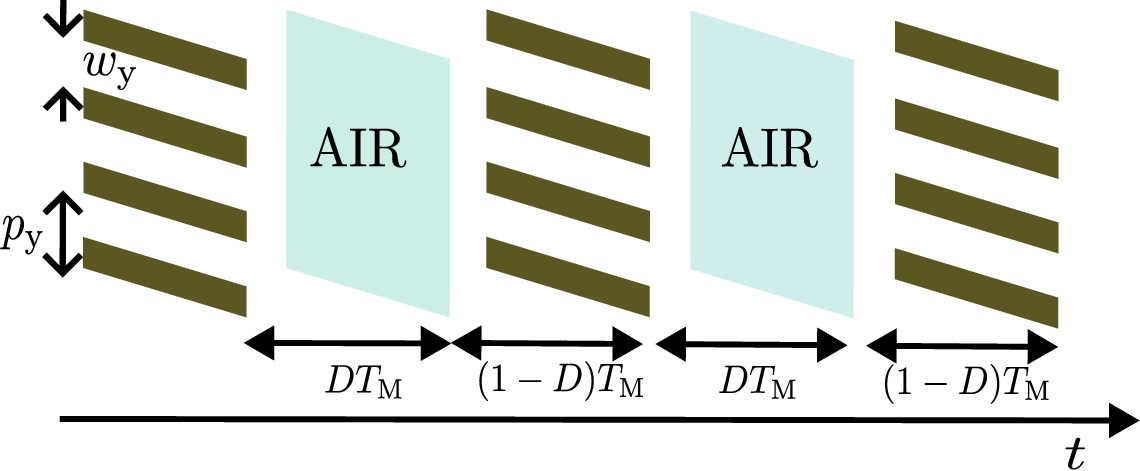} 
    \caption{Representation of the structure analyzed in \cite{Salva2024}. The structure is a time-modulated 1-D metallic grating. Time modulation is introduced by forcing the grating to become either transparent or reflective. Thus the structure keeps being a conventional grating during the time $D T_{\text{M}}$ and is transparent during $(1 - D)T_{\text{M}}$. This cycle is maintained indefinitely.}
    \label{grating}
\end{figure}

This scenario, involving negligible transient times, was studied and validated with an in-house FDTD code in \cite{Salva2024}. The structure analyzed in that paper consisted of an ideal grating \emph{instantaneously} switching between a conventional-grating state and fully transparent/reflective states along periodic cycles. Fig.~\ref{grating} shows an example in which the grating becomes transparent, and vice versa. The grating geometry is one-dimensional (1-D), so one of the spatial directions ($\hat{\mathbf{x}}$ in this case) is considered uniform from the electromagnetic point of view. 1-D systems represent simpler electromagnetic problems and can be readily reproduced using conventional in-house FDTD codes. Figs. 5-6 in \cite{Salva2024} validates the analytical model. The comparison of the harmonics' amplitudes obtained by the analytical model and FDTD shows excellent agreement. From these results, we can deduce that neglecting transient times or, in other words, assuming an instantaneous transition between metal and air (and vice versa), is a suitable approximation to adopt here.

\begin{figure}[t!] \includegraphics[width=0.7\columnwidth]{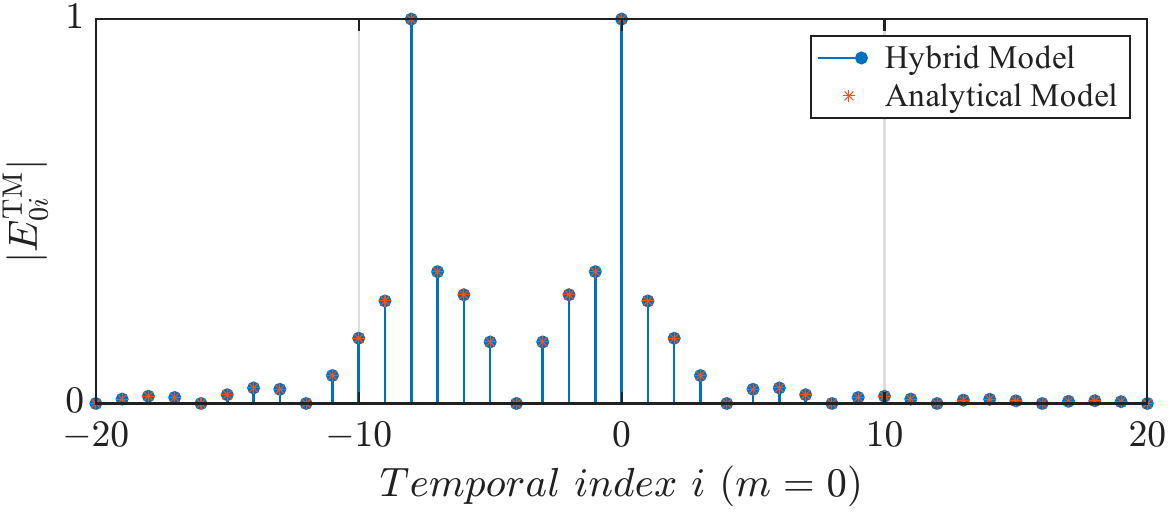}
    \caption{Amplitude estimation of the harmonics with $m = 0$. Both the analytical and hybrid approaches are employed. Structure parameters: $p = 10\,$mm, $w_{\text{y}} = p_{\text{y}}/2$, $T_{\text{M}} = 6T_{0}$.}
    \label{validation_1}
\end{figure}

A second partial validation comes from the hybridization of the macro function $\mathbf{E}_{\text{s}}(x, y, t)$. Hybridization consists of using analytical and numerical functions, together, to describe $\mathbf{E}_{\text{s}}(x, y, t)$. This is possible thanks to the factorization, 
\begin{equation}
\mathbf{E}_{\text{s}}(x, y, t) = \mathbf{E}(x, y)\cdot E(t)\,. 
\end{equation}
For non-canonical geometries, the spatial dependence $\mathbf{E}(x,y)$ can be obtained from commercial electromagnetic software and is therefore treated numerically \cite{Alex2021}. The temporal dependence, on the other hand, can be formulated analytically.

Let us now focus on the grating in Fig.~\ref{grating}. Under the assumption of negligible transient times, the evolution from one state to another is approximately regarded as \emph{instantaneous}. In a stationary state, the spatial profile is fundamentally dependent on the geometry of the scatterer and the polarization of the incident wave \cite{Berral2015}. The incident wave, in fact, determines the time evolution of the field profile, taking $E(t) = \sin(\omega_{0}t)$. 
In a time-modulated scenario, as that analyzed in \cite{Salva2024}, where the field at the discontinuity is described by the concatenation of two well-defined and distinct states, the macro function can be represented as
\begin{equation}\label{Eq_states}
\mathbf{E}(x, y, t) = A \sin(\omega_{0} t)\left\{
  \begin{array}{ll}
    \mathbf{E}_{1}(x, y), & 0 \leq t < D T_{\text{M}} \\
    \mathbf{E}_{2}(x, y), &  D T_{\text{M}} \leq t < T_{\text{M}},
  \end{array}
\right.
\end{equation}
with $\mathbf{E}_{1}(x, y)$ and $\mathbf{E}_{2}(x, y)$ being the entire-domain functions of each of the states, and
with $0 \le D \le 1$. The function $\sin(\omega_{0} t) $ introduces the time evolution of the macro function. In other words, the whole macro-basis function can be divided into a set of individual functions operating in a harmonic regime with time evolution given by $E(t) = \sin(\omega_{0}t)$. Of course, this can be assumed when the discontinuity plane of the unit cell is defined by scatterers that can be modeled via metals/air insertions. The introduction of lumped elements, for example, makes the scenario much more complex, and it would be beyond the approximations described here. The main challenge is to find scenarios where the scatterers of the unit cell can be represented by geometrical forms combining air and PEC. As discussed above, diodes may satisfy these requirements under the proper conditions.      

\begin{figure}[t!]
    \subfigure[]{\includegraphics[width=0.6\columnwidth]{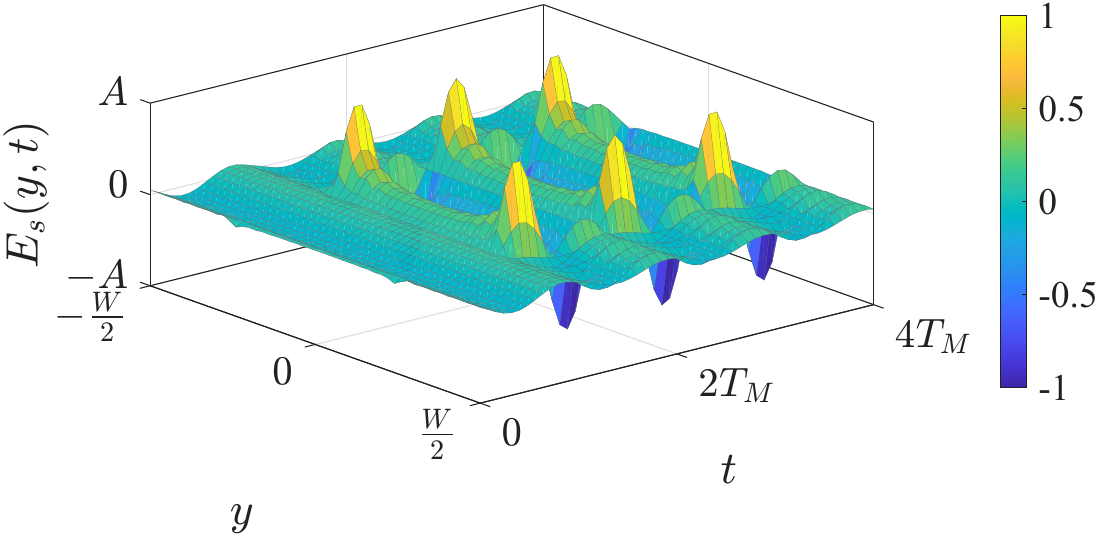}} \\
    \subfigure[]{\includegraphics[width=0.6\columnwidth]{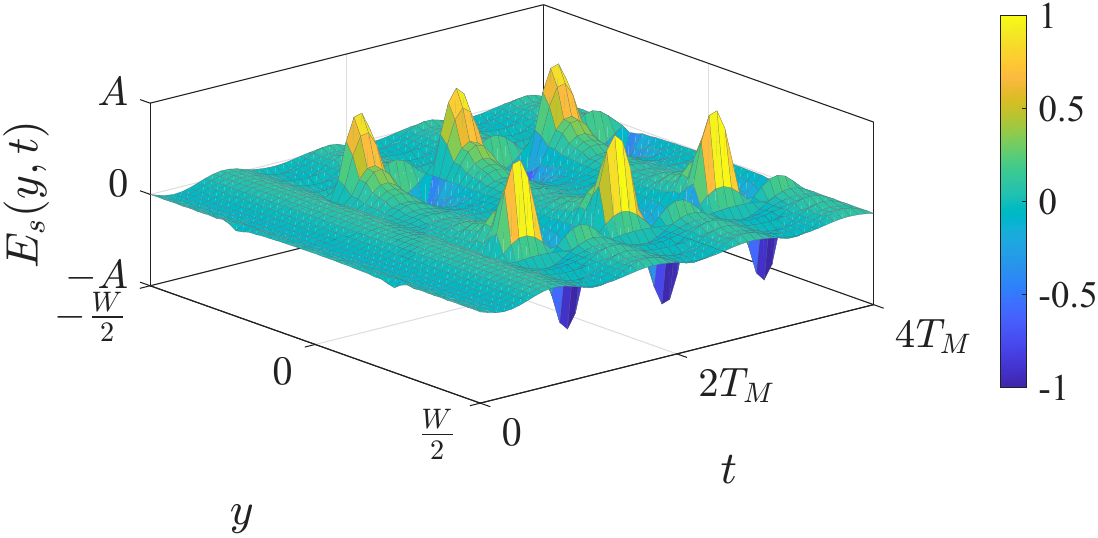}}
    \caption{(a): Representation of the electric field macro function  $\mathbf{E}_{\text{s}}(y, t)$ by using the analytical expressions. (c): Reconstruction of the electric field at $z = 0$ by the Floquet series. Structure parameters: $p = 10\,$mm, $w_{\text{y}} = p_{\text{y}}/2$, $T_{\text{M}} = 6T_{0}$.}
    \label{fig:Salva}
\end{figure} 

For the structure in Fig.~\ref{grating}, the macro basis function can be expressed as $\mathbf{E}(x, y, t) = \mathbf{E}(y) \cdot E(t)$, since $x$ is assumed spatially invariant. While the analytical function admits to be described as in \cite{Salva2024},
\begin{equation}
\mathbf{E}(y, t) = \hat{\mathbf{y}}\, A \sin(\omega_{0} t)  \left\{
\begin{array}{ll}
\frac{1}{\sqrt{1 - (2y/w_{\text{y}})^2}}, \hspace{5 mm} 0 \le t < DT_{\text{M}}   \\
1, \hspace{20.5 mm} DT_{\text{M}} \le t \le T_{\text{M}} \,,
\end{array}
\right.
\end{equation}
the hybrid function extracts the spatial profile $\mathbf{E}(y)$ from CST. The temporal evolution is $E(t) = \sin{\omega_{0}t}$ in both cases. 
Both versions of the macro function have been employed to estimate the amplitudes of a set of harmonics, by computing \eqref{E000cross}, \eqref{Enmico}, and \eqref{Enmicross}.
As shown in Fig.~\ref{validation_1}, we forced $m = 0$ and explored the evolution of the amplitudes with the temporal index $i$. The index $n$ is not necessary since the structure is periodic just along a single direction.
The excellent agreement between the results from the analytical and hybrid cases validates the possibility of using hybrid approaches, which are very useful for scatterers with complex and non-canonical shapes.

A third partial validation is a self-consistency test, in which the condition in \eqref{Econt1} should be satisfied. This condition establishes that at the discontinuity plane $(z = 0)$, the electric field described by the Floquet series is identical to the macro-basis function.  
Figs.~\ref{fig:Salva}(a)-(b) shows the field at the discontinuity plane along both $y$ and $t$. Fig.~\ref{fig:Salva}(a) represents the field obtained by the analytical macro function. Fig.~\ref{fig:Salva}(b) otherwise shows the reconstructed field obtained by the Floquet series of harmonics. The agreement is excellent, corroborating the model's self-consistency.

In summary, a set of partial validations has been conducted to assess the model's approximations. The resulting conclusions encourage us to employ the circuit approach to analyze two-dimensional structures such as those presented in the main text.